\documentclass[letterpaper, 10 pt, conference]{ieeeconf}  

\IEEEoverridecommandlockouts

\usepackage{graphicx} 
\usepackage{subcaption}
\usepackage{amsmath} 
\usepackage{mathtools}
\usepackage{amssymb}  
\usepackage{multicol}
\usepackage[T1]{fontenc}
\usepackage{algorithm}
\usepackage{algpseudocode}
\usepackage{hyperref}
\hypersetup{
    colorlinks=true, 
    linkcolor=blue,
    filecolor=magenta,      
    urlcolor=cyan,
    pdftitle={Overleaf Example},
    pdfpagemode=FullScreen,
    }

\usepackage{cite}
\usepackage{tabularx}
\usepackage{xcolor} 
\usepackage{soul}  
\usepackage{listings} 
\renewcommand{\vec}[1]{\boldsymbol{#1}} 
\newcommand{\mat}[1]{\vec{#1}}
\renewcommand{\bar}[1]{\overline{\mspace{-1mu}#1\mspace{1mu}}}
\newcommand{\mrm}[1]{\mathrm{#1}}
\newcommand{\mc}[1]{\mathcal{#1}}

\makeatletter
\renewcommand*\env@matrix[1][*\c@MaxMatrixCols c]{
  \hskip -\arraycolsep
  \let\@ifnextchar\new@ifnextchar
  \array{#1}}
\makeatother

\newcommand{\ssep}{\colon} 

\usepackage{xparse}     
\NewDocumentCommand{\widebar}{ O{0.8} O{2pt} m }{
    \mathrlap{\hspace{#2}\overline{\scalebox{#1}[1]{\phantom{\ensuremath{#3}}}}}\ensuremath{#3}
}
\NewDocumentCommand{\underwidebar}{ O{0.85} O{0.5pt} m }{
    \mathrlap{\hspace{#2}\underline{\scalebox{#1}[1]{\phantom{\ensuremath{#3}}}}}\ensuremath{#3}
}

\usepackage{stackengine}

\usepackage[normalem]{ulem} 
\usepackage{booktabs}
\usepackage{multirow}

\title{\LARGE \bf
Graph-Based Modeling, Control, and Optimization for \\
Multi-Domain and Multi-Timescale Energy Systems
}
\author{Joseph M. Pisani, Christopher T. Aksland, Philip M. Renkert, Joseph Broniszewski, Vismay Vyas, \\ \qquad Andrew G. Alleyne, Donald J. Docimo, Justin P. Koeln, Neera Jain, and Herschel C. Pangborn
\thanks{Joseph M. Pisani and Herschel C. Pangborn are with the Department of Mechanical Engineering, The Pennsylvania State University, University Park, PA 16802 USA (e-mail: {\tt\small hcpangborn@psu.edu}).}
\thanks{Christopher T. Aksland is with PC Krause \& Associates.} 
\thanks{Philip M. Renkert is with the University of Dayton Research Institute.}
\thanks{Joseph Broniszewski, Vismay Viyas, and Neera Jain are with the School of Mechanical Engineering, Purdue University.}
\thanks{Andrew G. Alleyne is with the College of Science and Engineering, University of Minnesota.}
\thanks{Donald J. Docimo is with the Department of Mechanical and Aerospace Engineering, Texas Tech University.}
\thanks{Justin P. Koeln is with the Department of Mechanical Engineering, University of Texas at Dallas.}
\thanks{This work was supported by the Office of Naval Research under Award N000142512051. Any opinions, findings, and conclusions or recommendations expressed in this material are those of the authors and do not necessarily reflect the views of the Office of Naval Research.}}

\begin{document}

\maketitle
\thispagestyle{empty}
\pagestyle{empty}

\begin{abstract}

Modern energy systems in vehicles and built infrastructure are governed by high-dimensional dynamics spanning multiple physical domains (e.g., electrical, thermal, mechanical) and timescales. This tutorial paper presents a graph-based modeling approach created to facilitate the modeling, analysis, control, estimation, optimization, and design of these systems. Matured and validated through more than a decade of research spanning multiple academic institutions and companies, the graph-based approach combines transient energy conservation with an explicit mathematical representation of the network by which energy is stored and transferred within a system. Following a mathematical overview of graph-based models, examples of multi-domain component and system models from the recent literature are presented, including single-phase thermal systems, two-phase thermal systems, and electro-mechanical systems. This is followed by a survey of recent applications for decentralized and hierarchical model predictive control, design optimization, and control co-design. Lastly, the paper describes an open-source toolbox created to facilitate the generation and analysis of graph-based models.

\end{abstract}

\section{Introduction} \label{sec:intro}

There is continual demand for energy systems in vehicles and built infrastructure to achieve greater performance, energy/power density, efficiency, sustainability, and resilience. This drives engineers towards systems with greater complexity, connectivity, and integration, in turn creating new opportunities and challenges in design, optimization, and control. The aviation field exemplifies this trend, where an exponential increase in electrical power on board advanced aircraft has contributed to a potential ``thermal management nightmare''~\cite{Walters2010}, inspiring calls to supplement conventional steady-state design processes with new methods that can exploit transients, as well as replace classical control architectures with more advanced frameworks~\cite{Doty2015b}. Similar calls can be found for virtually all transportation modalities, including but not limited to automobiles, trains, and naval ships~\cite{Alleyne2022}. The built environment is undergoing comparable trends~\cite{Wen2018}, spanning electrical power generation, storage, and distribution, as well as heating, ventilation, and air conditioning (HVAC). 

These applications often involve coupling across multiple physical domains. For example, electrified vehicle powertrains are necessarily electro-mechanical. Furthermore, electrical devices exhibit temperature-dependent reliability and must often be paired with thermal management systems. In turn, many thermal management systems are driven by electrical actuators and governed by thermal-fluid dynamics of single-phase or two-phase flows. Largely because of their multi-domain nature, the dynamics of these energy systems evolve across a wide range of timescales. For example, the temperatures of a large thermal energy storage module in a building might evolve over hours, while the voltages and currents of an electrical device can evolve over sub-millisecond timescales. This inherently leads to stiff dynamics. 

Technological innovations can be facilitated by access to dynamic models that are amenable to rapid simulation, symbolic analysis, scalable design optimization, and real-time model-based control. While data-based methods like machine learning offer many capabilities, there remain contexts in which physics-based models can offer greater insight and reliability by providing a first-principles, parametric representation of key optimization and control parameters, including physical properties, design parameters, and control inputs.

\subsection{Literature Review} \label{sec:literature}
A variety of established methods and tools exist for modeling complex energy systems. Prominent examples include Bond Graphs \cite{4140745} and Port-Hamiltonian Systems \cite{SYS-002}. These methods offer distinct advantages, such as the explicit enforcement of causality in Bond Graphs. The Modelica language \cite{1000174}, used in tools such as OpenModelica or Dymola, is an object-oriented modeling language designed for dynamic modeling of physical systems that then leverages a compiler for model simulation; many component libraries specifically for energy systems have been developed for academic and commercial use. Other commercial tools for modeling energy systems include Simscape, designed by MathWorks, that enables multi-domain modeling with MATLAB/Simulink integration \cite{MATLAB}.

A primary focus of these modeling frameworks is fidelity: faithfully reproducing the system's physical behavior in a simulation. However, to achieve the aforementioned technological innovations in energy systems requires models amenable to symbolic analysis, scalable design optimization, and real-time model-based control. An approach that offers this is graph-based modeling, or the expression of the system model as a network graph, composed of vertices and edges, that encodes the structure of energy conservation within a system. The first key benefit is deep system analysis. The compact graphical representation given by the network structure of a graph is not merely a schematic; \textit{it is a formal mathematical object}. This structure allows the rich toolkit of graph theory to be directly applied to analyze the model's internal interconnections. This enables powerful model decomposition techniques unavailable in many other frameworks. For example, large-scale models can be systematically partitioned into smaller, more manageable clusters using spectral graph tools to identify and separate components based on timescale interactions, or by using cut-set analysis to cluster subsystems based on the strength of their interactions. This decomposition is a powerful analytical tool, as it can algorithmically highlight specific topological features or hidden structures within the model. These revealed structures can then be directly capitalized upon by advanced controller designs, as demonstrated by the hierarchical control architectures later in this paper.

This analytical capability leads directly to the second major benefit: Inherent modularity and composability. The ability to decompose a system also implies the ability to compose one. Individual subsystems or components can be created, modeled, and validated in isolation. This ``re-composability'' is a key enabler for system design and optimization, creating a ``plug-and-play'' environment where subsystems can be easily interchanged to study various design tradeoffs and system configurations can be tested in a rapid, automated fashion. This flexibility, which allows design optimization to be performed concurrently with controller design, is far more streamlined in the graph-based setting than in monolithic modeling approaches.

The graph-based modeling approach described in this tutorial paper was first proposed in~\cite{Kolen_2016DSCC} for modeling single-phase fluid-thermal systems. Since 2016, it has been extended to a host of other components, subsystems, and systems, spanning two-phase refrigeration cycles, electro-mechanical powertrains, and turbomachinery. Table~\ref{tab:gbm_domains} summarizes the current graph-based modeling literature categorized by energy domains. As prior works include both single- and multi-domain energy systems, some studies appear in multiple rows. Table \ref{tab:gbm_applications} summarizes the applications of graph-based modeling for design, optimization, control, and system analysis.

\begin{table}[tb!]
    \centering
    \caption{Summary of graph-based model energy domains.}
    \label{tab:gbm_domains}
    \begin{tabularx}{\linewidth}{p{3cm}X}
        \toprule
        \textbf{Domain} & \textbf{References}\\  
        \midrule
        Electrical & \cite {Docimo_2020JMD, Thompson_2023CCTA, Williams_2017JDMC, Docimo_2018CCTA, Laird_2022Energy, Laird_2020DSCC, Aksland_2019ACC, Yu_2023Aviation, Koeln_202CST, Williams2015DSCC, Aksland_2021CEP, Aksland_2023CST, Yang_2018ATE, Docimo_2021JDMC, Belkacem_2025FKFS, Docimo2022ACC, Tannous2019_ACC} \\
        Mechanical & \cite {Bolander_2025SCITECH, Docimo_2020JMD, Thompson_2023CCTA, Williams_2017JDMC, Smith_2022ACC, Docimo_2018CCTA, Aksland_2019ACC, Yu_2023Aviation, Williams2015DSCC, Aksland_2021CEP, Aksland_2023CST, Yang_2018ATE, Docimo_2021JDMC, Belkacem_2025FKFS, Docimo2022ACC}\\
        Thermal & \cite {Aksland_2017DSCC, Bolander_2025SCITECH, Docimo_2020JMD, Bolander_2025TTE, Aksland_2025CSM, Thompson_2023CCTA, Echreshavi_2025CLIMA, Smith_2025ACC, Manion_2022iTherm, Williams_2017JDMC, Smith_2022ACC, Docimo_2018CCTA, Pangborn_2018JDMC, Kolen_2016DSCC, Laird_2022Energy, Laird_2020DSCC, Vyas_2026SCITECH, Russel_2022IJR, Aksland_2019ACC, Yu_2023Aviation, Pangborn_2017ACC, Lionello_2020Energy, Koeln_202CST, Williams2015DSCC, Pangborn_2020ACC, Aksland_2021CEP, Aksland_2023CST, Park_2023CCTA, Sisti_2025CLIMA, Yang_2018ATE, 10.1115/1.4043203, Docimo_2021JDMC, Belkacem_2025FKFS, Docimo2022ACC, ahu2026, Tannous2020_JDMC, Peddada2020_JDM, Tannous2019_CEP, Tannous2019_JDMC, Tannous2017_ACC, Peddada2017DETC, Tannous2020} \\
        \bottomrule
    \end{tabularx}
    \vspace*{-.5\baselineskip}
\end{table}

\begin{table}[tb!]
    \centering
    \caption{Applications of graph-based modeling.}
    \label{tab:gbm_applications}
    \begin{tabularx}{\linewidth}{p{3.95cm}X}
        \toprule
        \textbf{Application} & \textbf{References}\\  
        \midrule
        Architecture Optimization & \cite{Docimo_2020JMD, Bolander_2025TTE, Aksland_2025CSM, 10.1115/1.4043203, Belkacem_2025FKFS}\\
        Control Co-Design & \cite{Aksland_2025CSM, Thompson_2023CCTA, Smith_2025ACC, Laird_2022Energy, Laird_2020DSCC, Docimo_2021JDMC}\\
        Design Optimization & \cite{Docimo_2020JMD, Aksland_2025CSM, Yang_2018ATE, Belkacem_2025FKFS, Peddada2017DETC, Peddada2020_JDM}\\
        Estimation and/or Fault Detection & \cite{Tannous2019, shanks2023design, Tannous2020_JDMC, Tannous2019_CEP, Tannous2019_JDMC, Tannous2019_ACC, Tannous2017_ACC, Tannous2020} \\
        Model-Based Control & \cite{Pangborn_2018JDMC, Yu_2023Aviation, Pangborn_2017ACC, Koeln_202CST, Koeln2015DSCC, Williams2015DSCC, Pangborn_2020ACC, Aksland_2021CEP, Aksland_2023CST, Kolen_2018Automatica, Kolen_2017Automatica, Pangborn_2018ACC, Docimo2022ACC}\\
        System Analysis & \cite{Smith_2025ACC, Manion_2022iTherm, Yu_2023Aviation}\\
        \bottomrule
    \end{tabularx}
    \vspace*{-\baselineskip}
\end{table}

\subsection{Contribution and Outline}
In this tutorial paper, more than a decade of research on graph-based modeling for multi-domain energy systems is synthesized for the reader, beginning with a summary of the underlying mathematical description (Section~\ref{sec:mathoverview}) in both its original form and an extension to simultaneously encode multiple coupled energy conservation equations.  This is followed by examples of how graph-based modeling can be applied to represent single-phase and two-phase thermal-fluid components as well as electro-mechanical components (Section~\ref{sec:modelexamples}) and applications of the modeling approach for control analysis and design, control co-design, and design optimization (Section~\ref{sec:applications}). The paper also includes an overview of a new MATLAB-based toolbox developed to facilitate the creation and analysis of graph-based models (Section~\ref{sec:toolbox}). Lastly, Section~\ref{sec:conclusion} provides concluding remarks, discusses potential future directions, and identifies open problems.

\section{Mathematical Description} \label{sec:mathoverview}

This section first describes the original framework for graph-based modeling, applicable to a wide variety of mechanical, electrical, and single-phase thermal-fluid  components. For clarity of exposition, this is written under the assumption that the graph represents conservation of energy; however, as shown in Section~\ref{sec:single-phase}, the same general approach can be applied to represent conservation of mass. We then present an extension of the framework to simultaneously encode multiple coupled conservation equations, which later facilitates modeling of two-phase thermal systems in Section~\ref{sec:two-phase_example}. Lastly, we discuss how graph-based models of components can be connected to assemble system models. Unless otherwise stated, vectors and matrices are denoted by boldface letters, and sets by calligraphic letters. 

\subsection{Original Framework}
\label{sec:originalframework}
\begin{figure}[tb]
\centering
\includegraphics[width=0.85\columnwidth]{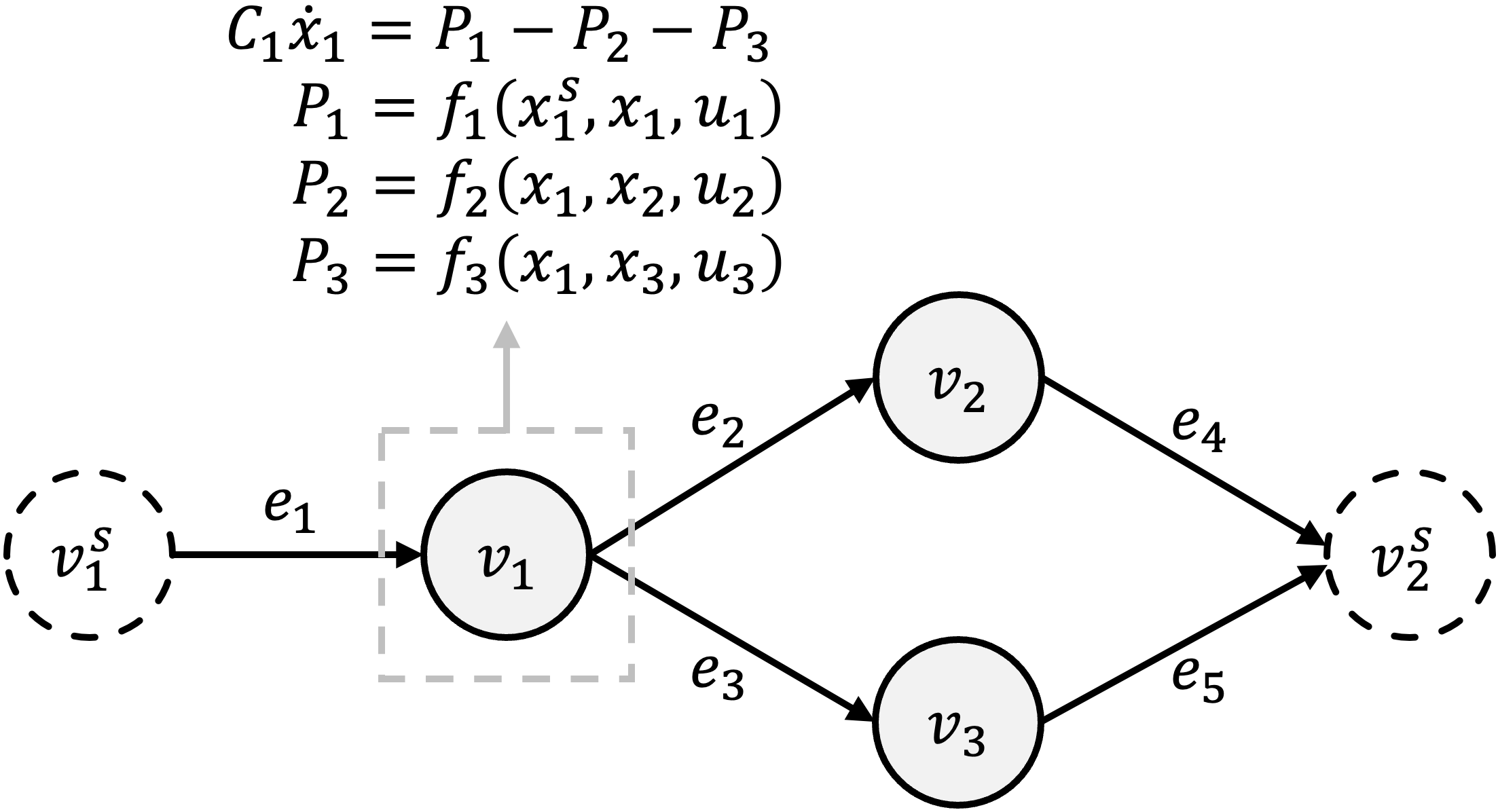}
 \caption{Notional example of a graph-based model.}
\label{fig:GBM Example}
\vspace*{-\baselineskip}
\end{figure}

Fig.~\ref{fig:GBM Example} illustrates a notional example of a graph-based model, represented as an oriented graph composed of vertices and edges. Vertices denote energy storage elements of the system while edges describe the energy transfer rates (referred to here as power flows) between interconnected energy storage elements. 
To define a graph-based model, let $\mc{V}$ be the set of vertices with cardinality $N_\mc{V}$ and $\mc{E}$ be the set of edges with cardinality $ N_\mc{E}$. 
Each edge $e_j \in \mc{E}$ has an orientation from its tail vertex to its head vertex, establishing the sign convention for positive power flow. In general, energy can be transferred in either direction along an edge, depending on the orientation of the edge and the sign of the associated power flow. 
The set of edges oriented out from vertex $v_{i}$ is denoted as $\mc{E}_{i}^{\mrm{tail}}$ 
and the set of edges oriented into $v_i$ is denoted as $\mc{E}_{i}^{\mrm{head}}$.

Vertices are classified into two categories: dynamic and external. Dynamic vertices are associated with dynamic energy storage states and are indicated by the vertex set $\mc{V}^{d} \subset \mc{V}$ with cardinality $N_{\mc{V}_{d}}$.
Each vertex in $\mc{V}^{d}$ therefore corresponds to a dynamic state $x_i(t)$. 
External vertices are associated with exogenous sources or sinks of power flow to the system, and are indicated by the vertex set $\mc{V}^{s} \subset \mc{V}$ with cardinality $N_{\mc{V}_{s}}$. Each vertex in $\mc{V}^{s}$ therefore corresponds to a time-varying disturbance to the system $x_i^s(t)$. Each edge corresponds to a power flow $ P_j(t)$. By conservation of energy, the differential equation corresponding to each dynamic state is given by
\begin{equation} \label{eq:GBM dynamics conservation law}
C_{i}\dot{x}_{i}(t) = \sum_{\{j \: \ssep e_{j} \in \mc{E}_{i}^{\mrm{head}}\}} P_{j}(t) \: - \sum_{\{j \: \ssep e_{j} \in \mc{E}_{i}^{\mrm{tail}}\}} P_{j}(t) \:,
\end{equation}
where $C_i$ is the energy storage capacitance corresponding to state $x_i$.

Each power flow is calculated as a function of the head and tail states of its corresponding edge, as well as a control input $u_j(t)$. In general, this can be expressed as a nonlinear function
\begin{equation} \label{eq:power flow}
\begin{split}
P_{j}(t) =f_j\left(x_{j}^{\mrm{tail}}(t),x_{j}^{\mrm{head}}(t),u_{j}(t)\right) \:.
\end{split}
\end{equation}

The graph structure is described by incidence matrix $\mat{M}\in \mathbb{R}^{N_\mc{V} \times N_{\mc{E}}}$ with
\begin{equation} \label{eq:GBM incidence matrix}
\begin{split}
&[m_{i,j}] =
\begin{cases}
\mspace{14mu} 1, & \text{if } v_{i} \in \mc{V} \text{ is the tail of } e_{j} \in \mc{E} \:, \\
-1, & \text{if } v_{i} \in \mc{V} \text{ is the head of } e_{j} \in \mc{E} \:, \\
\mspace{14mu} 0, & \text{otherwise}\:.
\end{cases}
\end{split}
\end{equation}
The incidence matrix can be partitioned according to dynamic and sink vertices as
\begin{equation} \label{eq:GBM incidence matrix partition}
\mat{M}=
\begin{bmatrix} \:
\widebar{\mat{M}} \: \: \\
\underwidebar{\mat{M}}
\end{bmatrix} \:,
\end{equation}
where $\widebar{\mat{M}} \in \mathbb{R}^{N_{\mc{V}_{d}} \times N_{\mc{E}}}$ and $\underwidebar{\mat{M}} \in \mathbb{R}^{N_{\mc{V}_{s}} \times N_{\mc{E}}}$. It is assumed that the vertices are ordered such that rows of $\widebar{\mat{M}}$ correspond to dynamic states and rows of $\underwidebar{\mat{M}}$ correspond to sink states.
A graph structure can be completely characterized by its vertices, edges, and incidence matrix as $\mc{G}=(\mc{V},\mc{E},\mat{M})$.

Based on the above notation, the system dynamics of a graph-based model are expressed as
\begin{equation} \label{eq:GBM dynamics}
\mat{C}\dot{\vec{x}}(t) = -\widebar{\mat{M}}\vec{P}(t)\:,
\end{equation}
where $\mat{C}$ is a diagonal matrix of the dynamic state capacitances, $\vec{x}$ is the vector of dynamic states, and $\vec{P}$ is the vector of power flows.

In some cases, it is convenient to also include external edges representing exogenous power flows to or from the system. Equation~\eqref{eq:GBM dynamics} then becomes
\begin{equation} \label{eq:GBM dynamics-DPin}
\mat{C}\dot{\vec{x}}(t) = -\widebar{\mat{M}}\vec{P}(t) + \mat{D}\vec{P}^{in}(t)\:,
\end{equation}

where $\vec{P}^{in}$ is the vector of external power flows and $\vec{D}$ captures the connectivity of external edges to dynamic vertices, similar to $\bar{\mat{M}}$.

In the remainder of the paper, we omit the dependence of states, power flows, and inputs on time except as necessary for clarity of exposition.

\subsection{Extension for Multi-State Vertices} \label{sec:two-phase_math}

As outlined above, graph-based models describe the system dynamics via conservation of energy. While energy conservation laws are practical for a wide range of energy system applications, models of systems with two-phase flow also require mass conservation. Additional state variables can be added to vertices of the graph to capture this. Russel et al. \cite{Russel_2022IJR} present a graph-based model with multi-state vertices, where pressure and enthalpy are chosen as the two dynamic states to describe a two-phase fluid system. The structure from \eqref{eq:GBM dynamics-DPin} is extended to encompass the multi-state framework, taking the form
 
\begin{equation}
    \mat{C}'\dot{\vec{x}} = -(\widebar{\mat{M}}*\bar{\mat{S}}_{\bar{\mat{M}}})\vec{P}(\vec{x},\vec{u}) + (\mat{D}*\mat{S}_{\mat{D}})\vec{P}^{in}\;,
\end{equation}
where the reader is referred to \cite{Russel_2022IJR} for the derivation and definition of terms. Bolander et al. \cite{Bolander_2025SCITECH} present an extension to \cite{Russel_2022IJR} using a combined multi-flow vector $\vec{\Gamma}$. The adaptive matrix $\vec{S}$ is used to map the multi-flow edges to the multi-state vertices in the graph model, where $\vec{S}$ is defined via the piece-wise expression
\begin{equation}\label{eq:GBM 2phase smatrix}
[s_{m,n}]_{i,j} =
\begin{cases}
1, & \text{if } (\mat{C} \dot{\vec{x}})_{i,m} = f(P_{j,n}) \\
  & \text{or if } (\mat{C} \dot{\vec{x}})_{i,m} = f(P_{s,j,n})\;, \\
0, & \text{otherwise}\;,
\end{cases}
\end{equation}
where $m$ is the state variable number of a given vertex $i$, $n$ is the flow edge type, and $i$ and $j$ correspond to vertices and edges of the incidence matrix, respectively.

A modified capacitance matrix $\mat{C}'$ extends the diagonal capacitance matrix $\mat{C}$ into a block diagonal capacitance matrix $\mat{C}'$, where block diagonal entries correspond to each vertex. A Khatri-Rao product $*$ between the incidence matrix  $\mat{M}$ and adaptive $\mat{S}$ matrix results in
\begin{equation} \label{eq:GBM 2phase dynamics}
{\mat{C}'}\dot{\vec{x}} = - (\mat{M} * \mat{S}) \vec{\Gamma} \;.
\end{equation}

\subsection{Connecting Graph-Based Models}\label{sec4_connecting}
Modeling complex and high-dimensional systems as graphs is often easiest by first building graph-based models of individual components and then connecting their vertices and edges. 
External vertices and edges denote a component's interaction with other components or the environment and are required to connect two (or more) components together.

\begin{figure}
    \centering
    \includegraphics[width=1\columnwidth]{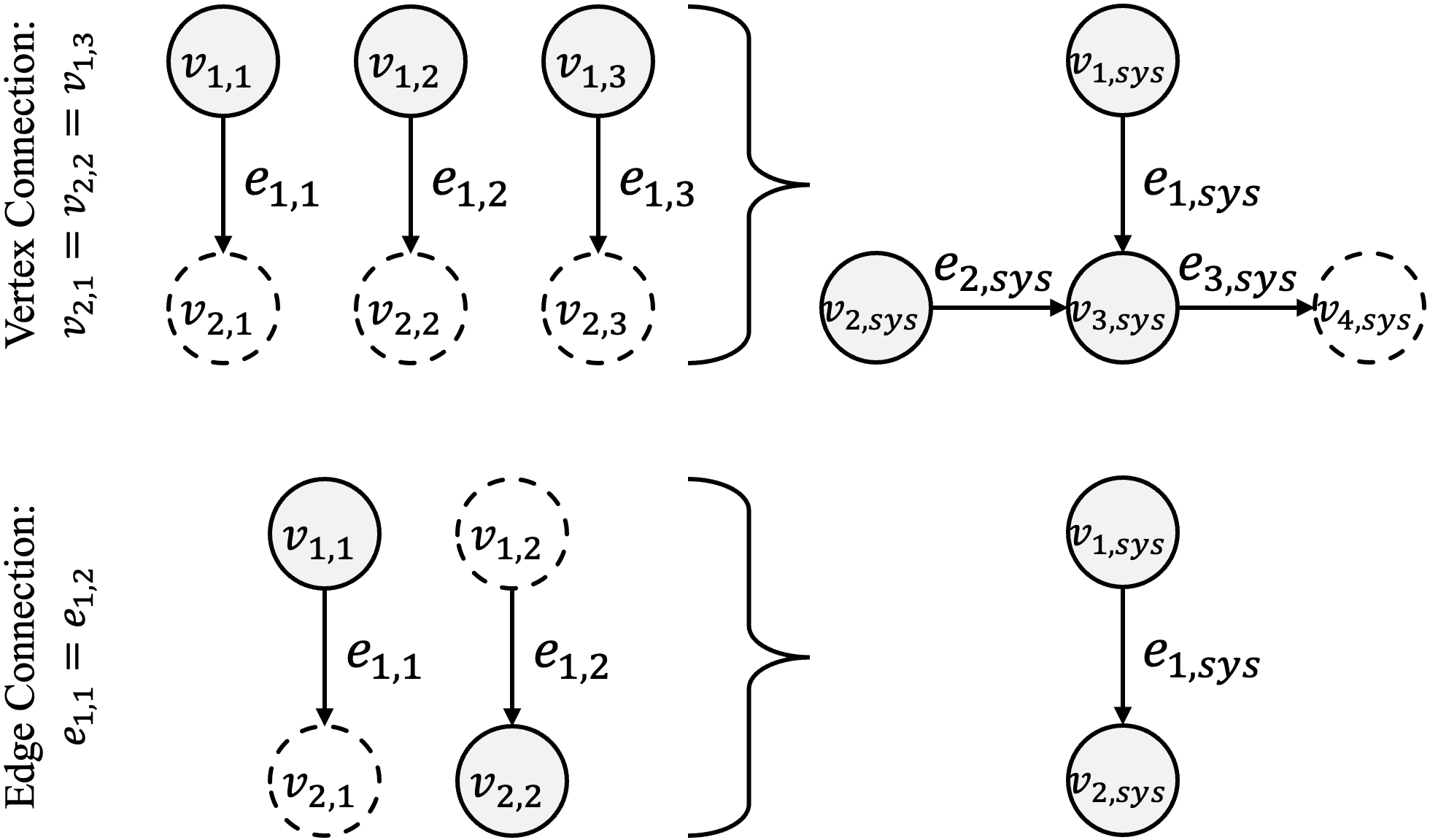}
    \caption{Visualization of vertex (top) and edge (bottom) connections, modified from \cite{Aksland_2019ACC}, where $x_{i,j}$ is the $i$-th state of component $j$.}
    \vspace*{-\baselineskip}
    \label{fig:ConnectionTypes}
\end{figure}

There are two ways to connect graph-based models: Vertex and edge connections, referred to as Type 1 and Type 2 connections, respectively, in \cite{Aksland_2019ACC}. Vertex connections are used when components have equivalent dynamic or external states, such as two electrical components in series that share the same current state or two components in parallel that share the same voltage state. The top of Fig.~\ref{fig:ConnectionTypes} shows how three simple graph-based models can be connected along a vertex. Note that a vertex connection can contain at most one dynamic vertex. 

When power is exchanged between two components, an edge connection is used. For example, the electrical power leaving a motor drive is equivalent to the power entering a connected electrical motor. A simple example is shown on the bottom of Fig.~\ref{fig:ConnectionTypes}.
\section{Model Examples} \label{sec:modelexamples}

With the mathematical framework of graph-based models now established, we next summarize how component- and subsystem-level models in several key physical domains can be derived via application of conservation of energy and mass. The toolbox described in Section~\ref{sec:toolbox} allows component or subsystem models to be coded \textit{once}, stored in a ``library,'' and then easily instantiated, parametrized, and interconnected as desired by the user. The existing toolbox library includes many models that been derived in the literature.

\begin{figure}[tb]
\centering
\includegraphics[width=\columnwidth]{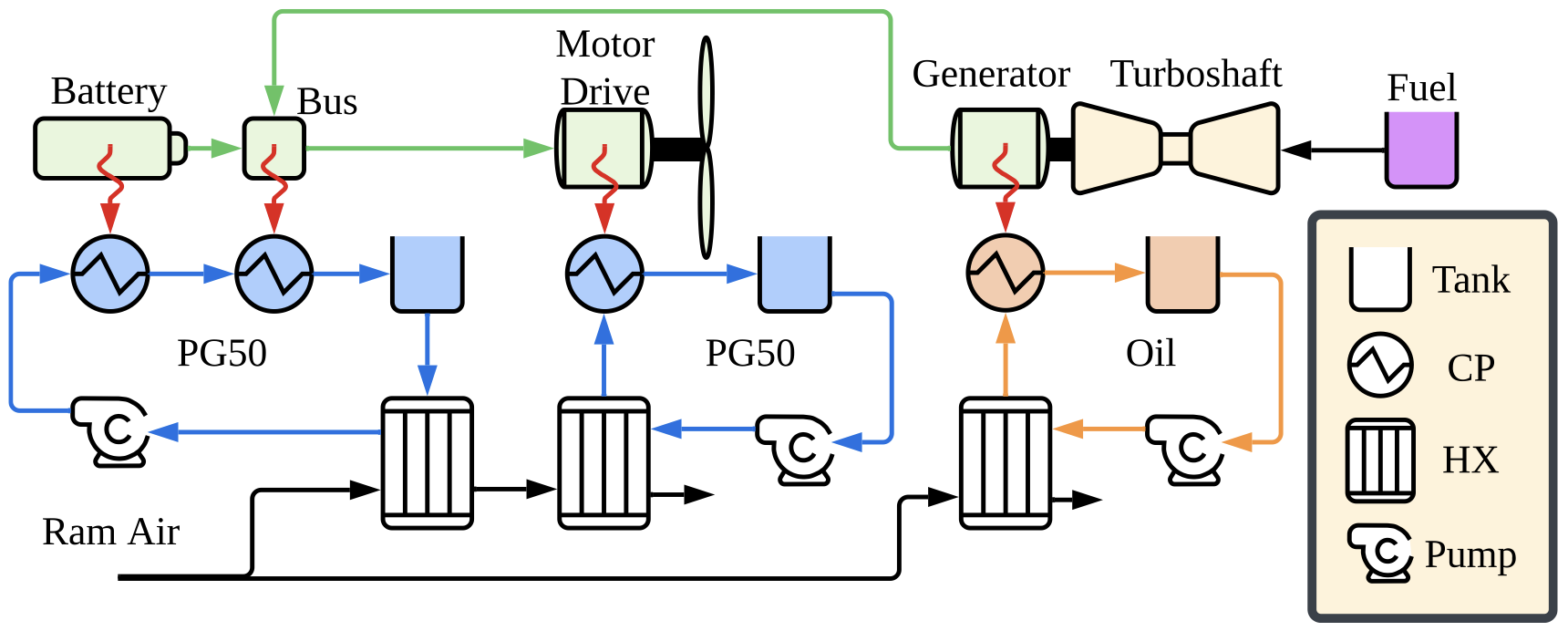}
\includegraphics[width=\columnwidth]{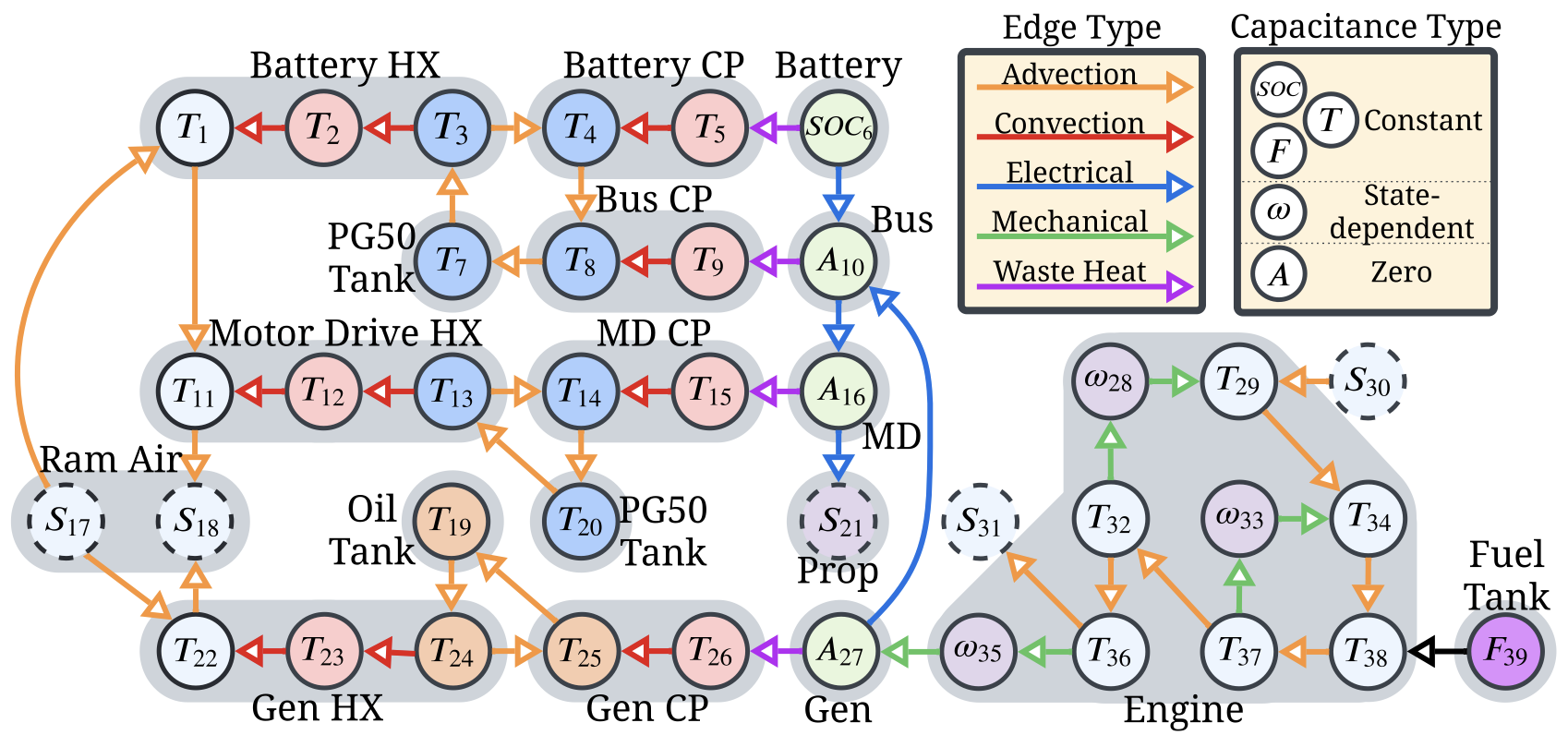}
 \caption{Schematic (top) and corresponding graph-based model structure (bottom) for a hybrid-electric aircraft propulsion, power, and thermal management system \cite{Thompson_2023CCTA}.}
 \vspace*{-\baselineskip}
\label{fig:Graph-example}
\end{figure}

\begin{figure}[tb]
\centering
\includegraphics[width=\columnwidth]{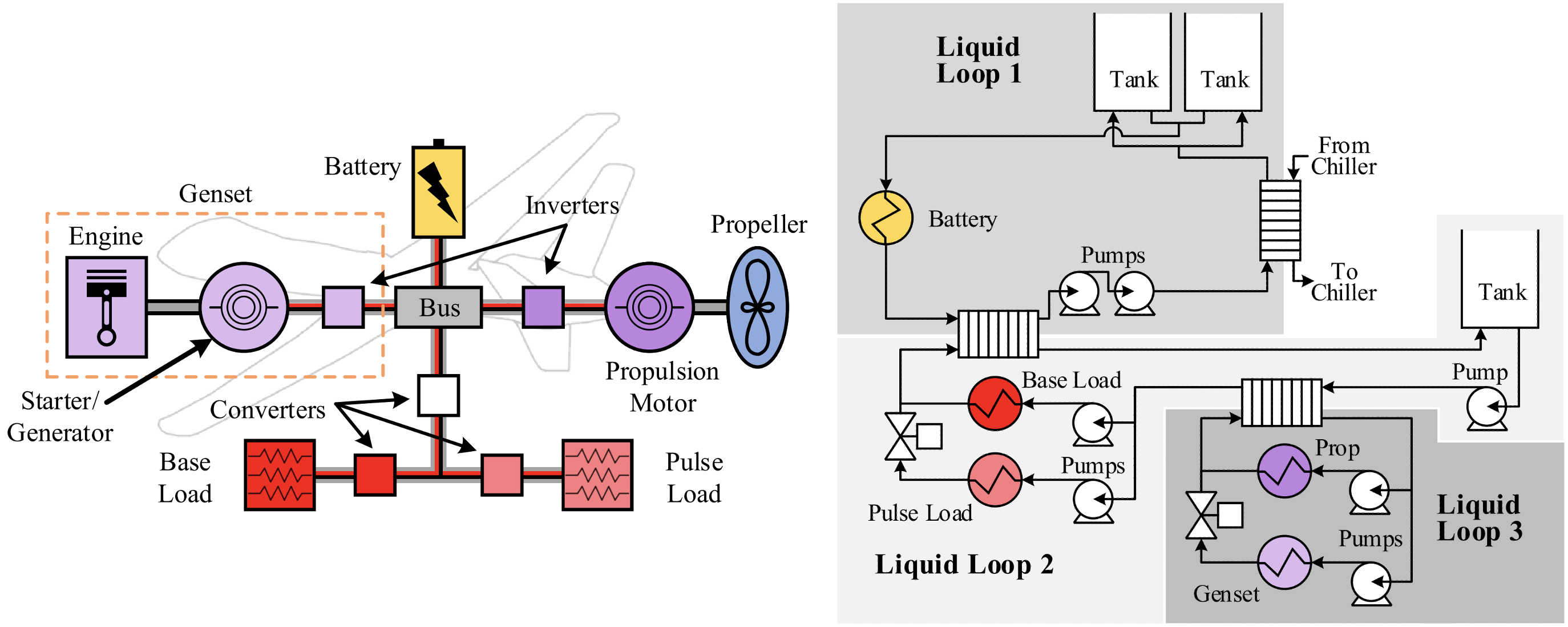}

\vspace{3mm}
\includegraphics[width=\columnwidth]{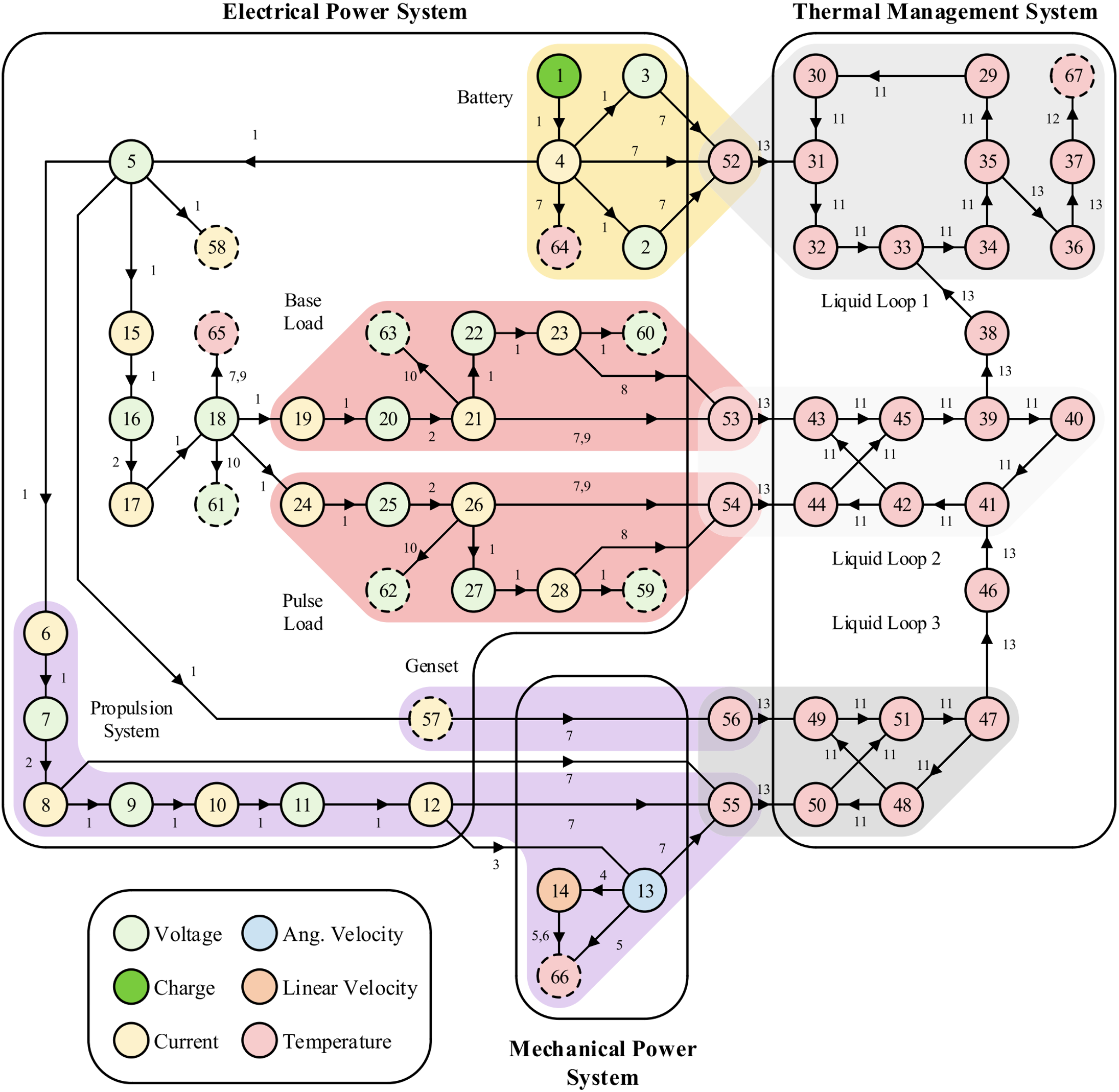}
 \caption{Schematic (top) and corresponding graph-based model structure (bottom) for a UAV powertrain and thermal management system~\cite{Aksland_2023CST}.}
 \vspace*{-\baselineskip}
\label{fig:Graph-example2}
\end{figure}

While the discussion in this section is organized by physical domain, a key feature of graph-based modeling is the ability to interconnect component models spanning different domains to represent complex multi-domain systems. 
As an example, Fig.~\ref{fig:Graph-example} from~\cite{Thompson_2023CCTA} shows a candidate architecture for propulsion, power, and thermal management systems of a hybrid-electric aircraft (top) and the corresponding graph-based model structure (bottom), formed by coupling individual graph-based models for each key component and subsystem. This uses a simplified representation for electrical components in terms of average power, whereas Section~\ref{sec:electrical-modeling} presents a higher-fidelity approach that captures voltage and current transients. Fig.~\ref{fig:Graph-example2} shows a second example from~\cite{Aksland_2023CST} for an uncrewed aerial vehicle (UAV) powertrain and thermal management system that uses these higher-fidelity electrical component representations.

\subsection{Single-Phase Thermal and Fluid Systems} \label{sec:single-phase}

The reader is referred to~\cite{Kolen_2016DSCC} for a detailed derivation of graph-based models for single-phase thermal components. For these components, conservation of thermal energy is applied within the graph-based framework to yield vertex states representing temperatures of lumped thermal control volumes $T$ and edge power flows representing heat transfer between control volumes. The control inputs to the model are typically fluid mass flow rates $\dot{m}$.

For all single-phase thermal vertices, the capacitance is given by $ C =M c_p $, where $M$ is the control volume mass and $ c_p $ is the specific heat capacitance of the control volume fluid or material, which can either be assumed constant or modeled as a function of temperature.
For advective heat transfer due to fluid flow, the power flow equation~\eqref{eq:power flow} specifies to
\begin{equation}
P= \dot{m} c_p T^{\text{tail}} \;,
\end{equation}
where we assume that fluid flows in the same direction as the orientation of the edge. That is, the rate of advective heat transfer is equal to product of the mass flow rate input, specific heat capacitance, and upstream temperature state. 
For convective heat transfer in heat exchangers, the power flow is
\begin{equation}
P=hA_{s}(T^{\text{tail}}-T^{\text{head}}) \;,
\end{equation}
where $ A_s $ is the convective surface area and $ h $ is the heat transfer coefficient, which can either be assumed constant or modeled as a function of the head temperature, tail temperature, and fluid mass flow rate.

While in many cases it is sufficient to treat the pressure dynamics of single-phase systems as operating at pseudo-steady state relative to the timescale of the thermal dynamics, it is also possible to apply conservation of mass in the graph-based framework to capture pressure dynamics. In this case, vertex states represent the pressures of lumped control volumes $p$ and edges represent mass flow rates between control volumes. The inputs to the model are typically actuator states such as the rotational speed of centrifugal pumps or percentage opening of a valve orifice.

For all pressure vertices except those of a reservoir, the hydraulic capacitance is given by $ C= M/E $, where $ E $ is the bulk modulus of the fluid.  For reservoirs, $ C= A_{c}/g $, where $ A_c $ is the reservoir cross sectional area and $ g $ is the gravitational constant. 
For most hydraulic edges (e.g., in cold plates and heat exchangers), the edge equation specifies to
\begin{equation}
\dot{m}=\rho A_{c} \sqrt{\frac{2(p^{\text{tail}}-p^{\text{head}}+\rho g \Delta h)}{\rho (f \frac{L}{D} +K_{L})}} \;,
\label{eq:HX_mfr}
\end{equation}
where $P^{\text{tail}}$ and $P^{\text{head}}$ are the upstream and downstream pressure states, respectively, $ L $, $ D $, and $ A_c $ are the fluid flow length, diameter, and cross sectional area, respectively, $ \Delta h $ is the height difference between the inlet and outlet flow, $ f $ is the friction factor, and $ K_L $ is the minor loss coefficient. For centrifugal pumps, the mass flow rate is given by
\begin{equation}
\dot{m}=\rho A_{c} \sqrt{2g \bigg( H-\frac{p^{\text{head}}-p^{\text{tail}}}{\rho g} \bigg) } \;,
\end{equation}
where the pump head $ H $ is determined using an empirical map, typically as a function of pump rotational speed $ \omega $ and the pressure difference across the pump. This again assumes that fluid flows in the same direction as the orientation of the edge.

\begin{figure}
    \centering
    \includegraphics[width=.8\columnwidth]{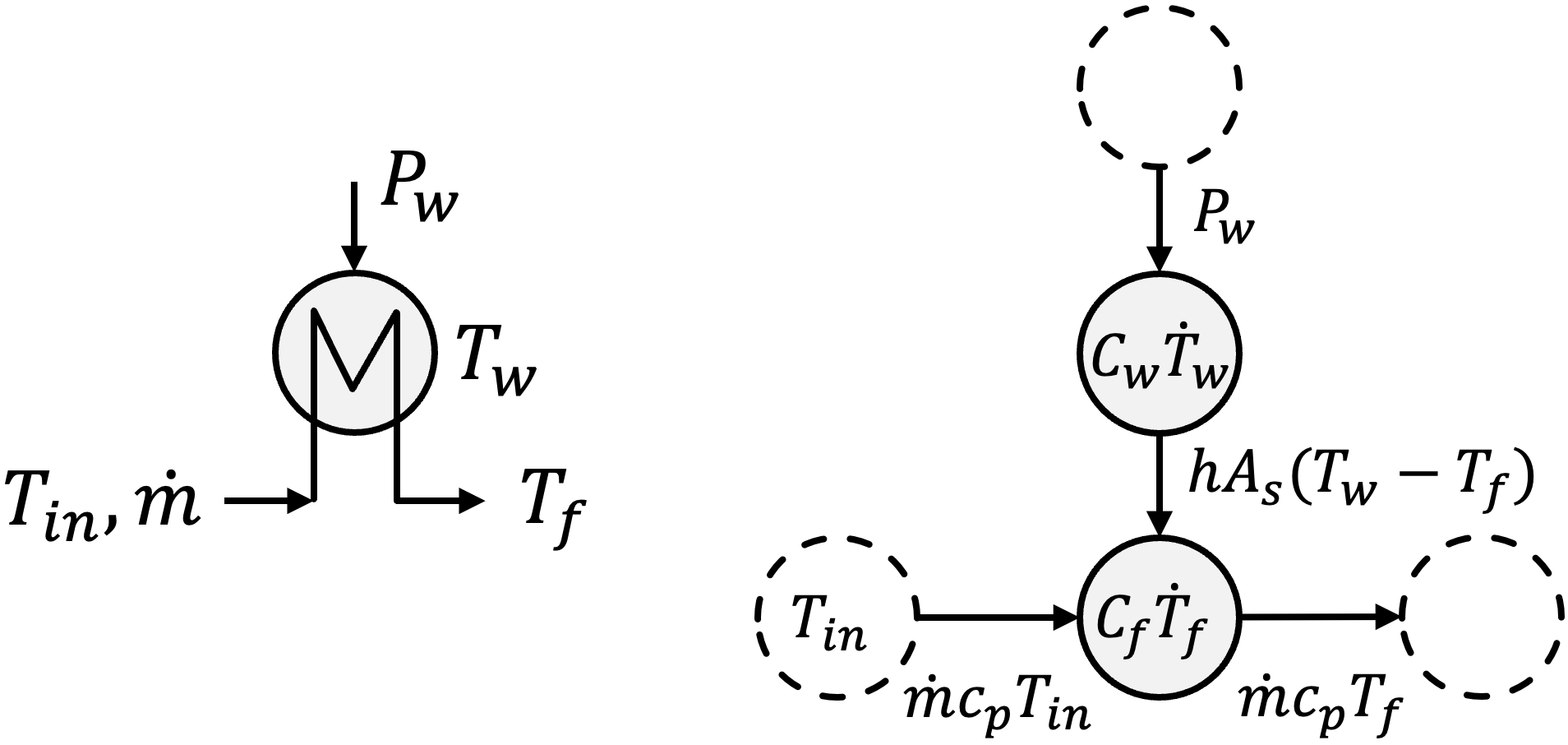}
    \caption{Example graph-based model for a cold plate with single-phase flow.}
    \vspace*{-\baselineskip}
    \label{fig:SinglePhaseGraph}
\end{figure}

As a representative example, Fig.~\ref{fig:SinglePhaseGraph} illustrates the thermal graph-based model of a single-phase cold plate heat exchanger assuming that pressure dynamics operate at pseudo-steady state, where subscripts $in$, $f$, and $w$ denote the upstream fluid, lumped fluid in the cold plate, and lumped cold plate wall, respectively, and $P_w$ is the heat transfer into the wall from a device cooled by the cold plate.

When both temperature and pressure dynamics of single-phase components are to be simulated together, the fluid mass flow rates computed as edge properties of the pressure graph can be communicated as edge inputs to the thermal graph. In turn, fluid properties of the pressure graph can be calculated as a function of the temperature states of the thermal graph. In this way, the two graphs are bidirectionally coupled but can be solved in parallel at each numerical integration time step and without any algebraic loops. That is not the case for two-phase systems, as alluded to in Section~\ref{sec:two-phase_math} and described in further detail next.

\subsection{Two-Phase Thermal and Fluid Systems} \label{sec:two-phase_example}

\textcolor{black}{
In two-phase fluid systems, single-state vertices are insufficient to determine thermodynamic states due to the strong thermodynamic coupling between pressure $p$ and enthalpy $h$ in the two-phase region. The reader is referred to \cite{Russel_2022IJR} for a detailed derivation of a graph-based model for a two-phase thermal component. Here we summarize the derivation of a multi-state heat exchanger model. 
}
Applying conservation of mass and energy to a two-phase heat exchanger yields
\begin{align}
\dot M_{r}&=\frac{d(\rho_{r} V_r)}{dt}=\dot{m}_{in} -\dot{m}_{out}\;,
\label{eq:two_phase_mass} \\
\dot E_{r}&=\frac{d(\rho_{r} h_{r} V_r)}{dt}=\dot{m}_{in} h_{in}-\dot{m}_{out} h_{out}+P_r\;,
\label{eq:two_phase_energy}
\end{align}
where the subscript $r$ refers to refrigerant and $P_r$ is the convective heat transfer between the wall and refrigerant. 
Expanding $\dot M_{r}$ and $\dot E_{r}$ results in
\begin{align}
\left(
\left.\frac{\partial \rho_{r}}{\partial p_{r}}\right|_{h_r} \dot{p}_{r} + \left.\frac{\partial \rho_{r}}{\partial h_{r}}\right|_{p_r} \dot{h}_{r} \right)V_r+\dot{m}_{out}-\dot{m}_{in}&=0\;,
\label{eq:two_phase_mass_final}\\
\Phi_1\dot p_{r}+\Phi_2\dot h_{r}+\dot{m}_{out} h_{r,out} -\dot{m}_{in} h_{r,in}-P_r&=0\;,
\label{eq:two_phase_energy_final}
\end{align}
where
\begin{align}
\Phi_1&=\left.\left(\frac{\partial \rho_r}{\partial p_r}\right|_{h_r} h_r-1\right) V_r\;, \\
\Phi_2&=\left(\left.\frac{\partial \rho_r}{\partial h_r}\right|_{p_r} h_r+\rho_r\right) V_r\;.
\end{align}
The quantities appearing in $\Phi_1$ and $\Phi_2$, including the density $\rho$ and its thermodynamic derivatives $\left.\left(\partial \rho_{r} / \partial p_{r}\right)\right|_{h_r}$ and $\left.\left(\partial \rho_{r} / \partial h_{r}\right)\right|_{p_r}$, are obtained using fluid property lookup tables. Similarly, other properties such as temperature and vapor quality must generally be evaluated as functions of the local state. The heat transfer coefficient appearing in the power flow describing convective heat transfer, $ {P_r}=hA_s(T^{tail}-T^{head}) $, may vary significantly with phase and flow regime and is typically modeled using empirical correlations. The mass flow rates are given by \eqref{eq:HX_mfr}.

\begin{figure}
    \centering
    \includegraphics[width=.8\columnwidth]{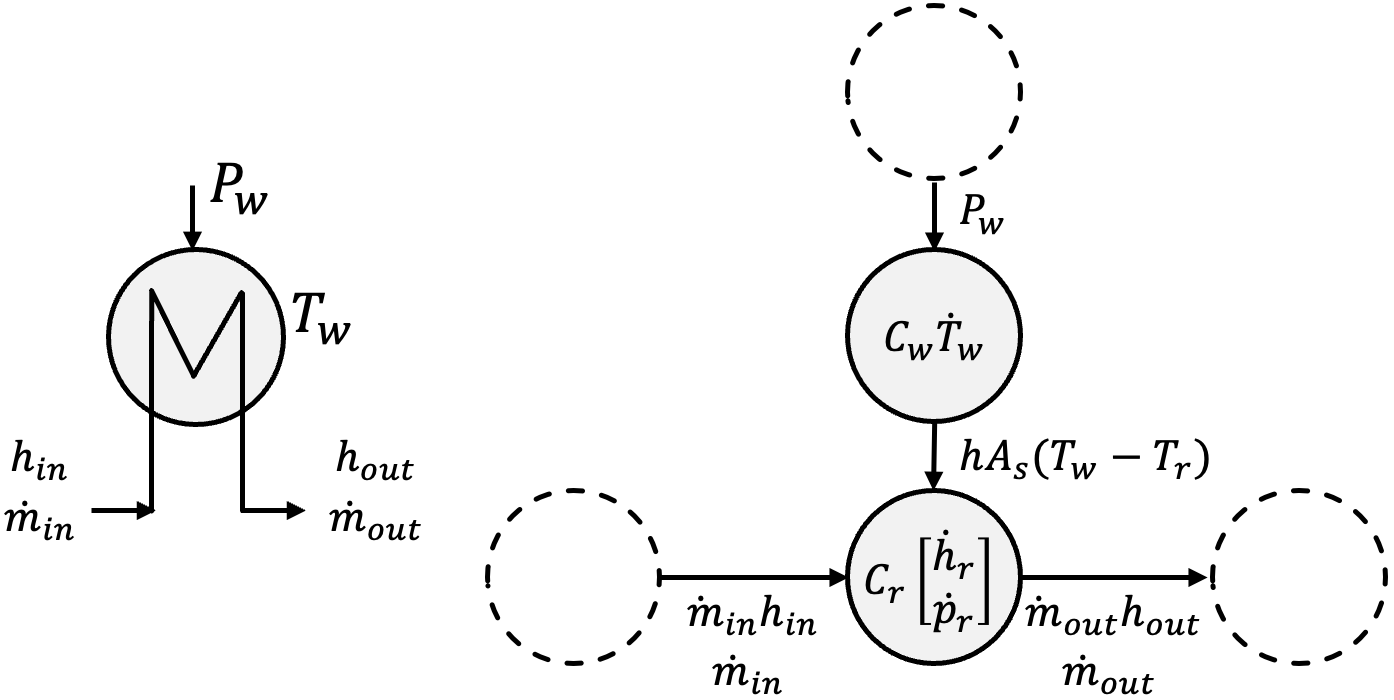}
    \caption{Example graph-based model for a cold plate with two-phase flow.}
    \label{fig:TwoPhaseGraph}
\end{figure}

As a representative example, Fig.~\ref{fig:TwoPhaseGraph} illustrates the graph-based representation of a two-phase fluid cold plate heat exchanger formulated using a two-state vertex. The state vector is defined as
\begin{equation}
\vec{x} =
\begin{bmatrix}
h_r & p_r & T_w
\end{bmatrix}^T,
\end{equation}
and the flow vector is defined as
\begin{equation}    
\vec{\Gamma} =
\begin{bmatrix}
\dot{m}_{in} h_{in} & \dot{m}_{in} & 
\dot{m}_{out} h_r & \dot{m}_{out} &
P_r & P_{w}
\end{bmatrix}^T \;.
\end{equation}
From \cite{Russel_2022IJR}, the capacitance matrix for the system is defined as
\begin{equation}
\mat{C}' =
\begin{bmatrix}
\mat{C}_r & 0\\ 0 & C_w
\end{bmatrix}\;,
\end{equation}
where
\begin{align}
\mat{C}_r &=
\begin{bmatrix}
\Phi_2 & \Phi_1 \\[1.2em]
\left.\frac{\partial \rho_r}{\partial h_r}\right|_{p_r} V_r & \left.\frac{\partial \rho_r}{\partial p_r}\right|_{h_r} V_r
\end{bmatrix}\;, \\
C_w &= (M c_p)_{w}\;.
\end{align}
From \cite{Bolander_2025SCITECH}, the full system can be described by~\eqref{eq:GBM 2phase dynamics} where 
\begin{equation}
\mat{M}*\mat{S}=
\begin{bmatrix}
-1 & 0 & 1 & 0 & -1 & 0\\
0 & -1 & 0 & 1 & 0 & 0\\
0 & 0 & 0 & 0 & 1 & -1
\end{bmatrix}\;.
\end{equation}

Unlike in this simplified example, which uses a single control volume, two-phase heat exchangers are typically discretized into multiple control volumes to capture spatial variations in fluid states.
  
The single-phase thermal-fluid system formulation described in Section~\ref{sec:single-phase} can be viewed as a special case of the two-phase thermal-fluid system formulation in which phase change is absent. In these cases, the thermodynamic process can be adequately described by a single state, reducing to the simpler single-state representation.

The multi-state vertex graph formulation can be applied to other thermal-fluid components and other combinations of thermodynamic states. An example of such an application is presented in \cite{Bolander_2025SCITECH} for modeling turbomachinery, in which air is the working fluid and the relevant thermodynamics are dependent on both temperature and pressure dynamics.

\subsection{Electrical and Mechanical Systems} \label{sec:electrical-modeling}
Instead of presenting electrical and mechanical system modeling separately, they are presented together in the context of the mechanical-electrical analogy, the concept that mechanical systems can be represented as electrical networks and vice versa. Table \ref{tab:EMAnalogy} provides relevant similarities in the mechanical-electrical analogy that should be referenced while reading this section. Only translational mechanics are presented here, but the concepts can be extended to rotational mechanics as well.

\begin{table}[tb]
\centering
\caption{Analogous modeling elements for mechanical and electrical systems.}
\label{tab:EMAnalogy}
\begin{tabular}{@{}cccc@{}}
\toprule
                            &             & Compliant        & Inertial               \\ \cmidrule(l){2-4} 
\multirow{2}{*}{Mechanical} & State       & Force $F$        & Velocity $v$           \\
                            & Capacitance & Displacement $x$ & Momentum $p$           \\ \midrule
\multirow{2}{*}{Electrical} & State       & Voltage $V$      & Current $I$            \\
                            & Capacitance & Charge $Q$       & Flux Linkage $\lambda$ \\ \bottomrule
\end{tabular}
\vspace*{-\baselineskip}
\end{table}

Both energy domains store energy in compliant  (springs and capacitors) and inertial (masses and inductors) elements that correspond to force (spring), voltage (capacitor), velocity (mass), and current (inductor) states for the vertices of a graph-based model. Subsequently, the capacitance for each vertex is displacement $d$, charge $Q$, momentum $p$, and flux linkage $\lambda$. For many elements, the capacitance is the product of the element's state variable $x$ and characteristic property $\lambda_p$ as
\begin{equation}
    C=\lambda_p x \;.
\end{equation}
For fundamental mechanical and electrical components, the characteristic properties are spring compliance, electrical capacitance, mass, and inductance. In some cases it may be more convenient to represent the state and capacitance in other ways, such as an electrical battery where the state and capacitance are the state of charge and energy capacity, respectively. The capacitance of each element may be a nonlinear function of the state.

In electrical and mechanical systems, there are a variety of ways that power can be transmitted or transformed, many more than can be exhaustively documented in this paper. Therefore, only the most common are described here. First, conventional mechanical and electrical power is useful for representing power transmission in a graph-based model. In this case, power is the product of the state variables: $P=Fv$ for mechanical systems and $P=VI$ for electrical systems, or more generally
\begin{equation}
    P=x^{\text{head}}x^{\text{tail}} \;.
\end{equation}
Since power flows can be a function of at most two states, this implies that power transmission must be between alternating compliance and inertia elements (e.g., a force vertex adjacent to a velocity vertex). When building a graph-based model from a schematic, this concept may be counter-intuitive because it is common to find, for example, two capacitors in series. Such elements are directly connected in the circuit, but may not be directly connected in the graph-based model.

There are two equivalent methods to address this challenge. One approach is to simplify the system by combining the elements into a single lumped element (e.g., combine capacitors in series into a single equivalent capacitance). While this approach works well for many systems, there are some cases where the system is difficult or non-intuitive to simplify, like non-reducible topologies with algebraically coupled energy storage elements (e.g., an ideal transformer with capacitors across the primary and secondary windings). In these cases, one can insert a \textit{virtual} element between the two elements (e.g., insert an inductor in series between the two capacitors. Virtual elements have zero capacitance, creating an algebraic coupling between their adjacent elements. Placing virtual elements in a graph-based model is useful when building increasingly complex systems. Due to the zeroed capacitances, the differential equations in~\eqref{eq:GBM dynamics} and \eqref{eq:GBM dynamics-DPin} then become differential-algebraic equations (DAEs).

Another common power flow to represent is energy loss, often modeled using dampers or resistors in mechanical and electrical systems. These terms are useful to model the coupling between electrical and mechanical systems since mechanisms such as friction in gears or switching losses in power converters generate heat. Energy loss can generally modeled as
\begin{equation}
    P= \lambda_p(x^{\text{tail}})^2 \;,
\end{equation}
where $\lambda_p$ is the element's characteristic property like the dampening coefficient or resistance value (or their inverses) and may be a nonlinear function and/or input-dependent. While this is one common approach to model energy loss, there are variety of other ways that energy loss terms can be represented, such as with static friction $P=\lambda_px^{\text{tail}}\text{sign}\left(x^{\text{tail}}\right)$, where $x^{\text{tail}}$ is a velocity vertex. Using these forms, energy transfer is unidirectional, always flowing out of the electrical or mechanical elements.

A third common power flow representation is power conversion, often present in devices such as gears, motors, power converters, etc. Power conversion can commonly be represented as
\begin{equation}
    P=\lambda_px^{\text{head}}x^{\text{tail}} \;,
\end{equation}
where $\lambda_p$ is the element's conversion ratio like the gear ratio or transformer ratio and may be a nonlinear function and/or input-dependent (e.g., a variable-speed transmission). 

\begin{figure}[tb]
    \centering
    \begin{subfigure}[b]{0.5\textwidth}
    \centering
    \includegraphics[width=.53\linewidth]{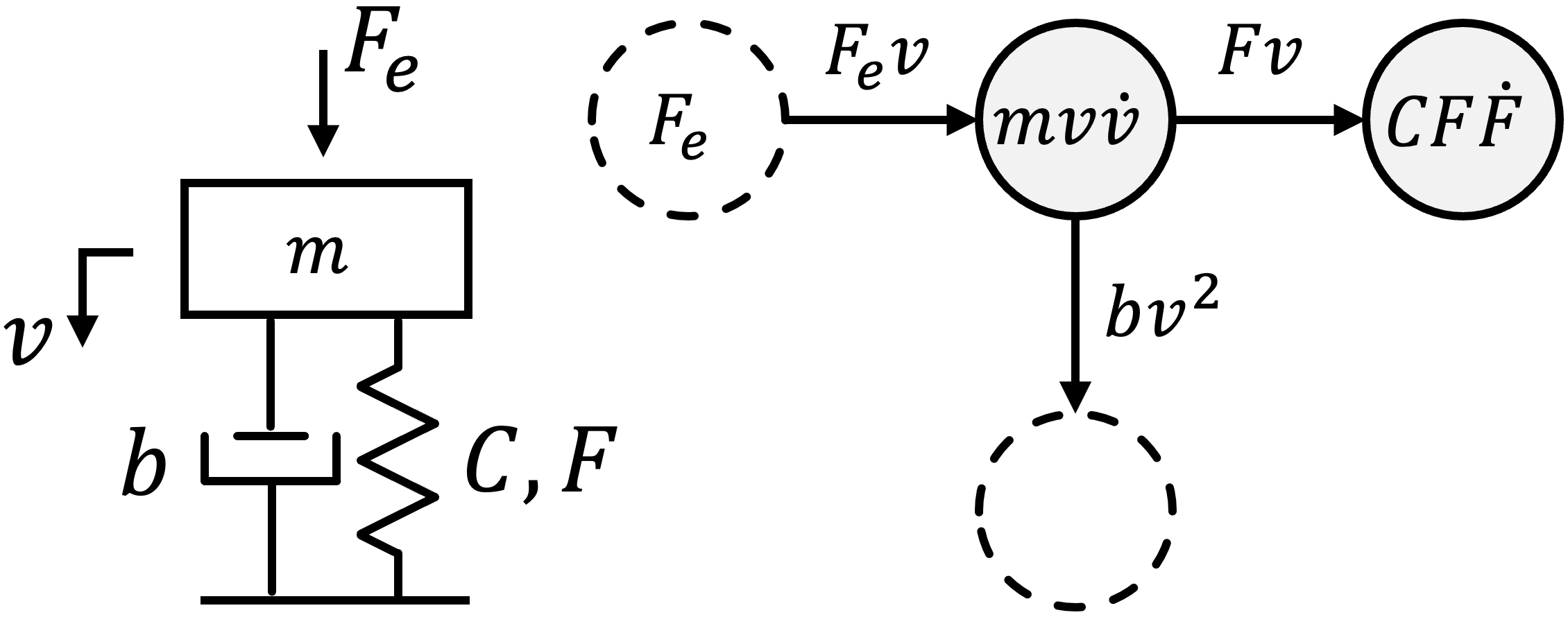}
        \caption{Mass-spring-damper}
        \label{fig:sub1}
    \end{subfigure}
    \hfill 
    \begin{subfigure}[b]{0.5\textwidth}
    \centering

    \includegraphics[width=1\linewidth]{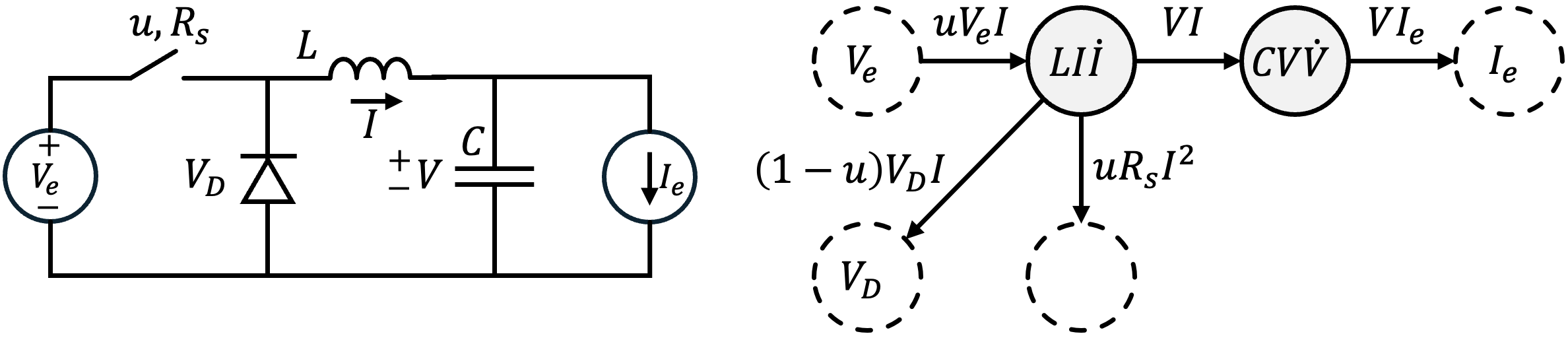}
        \caption{Buck-converter}
        \label{fig:sub2}
    \end{subfigure}
    \hfill 
    \begin{subfigure}[b]{0.5\textwidth}
    \centering
    \includegraphics[width=1\linewidth]{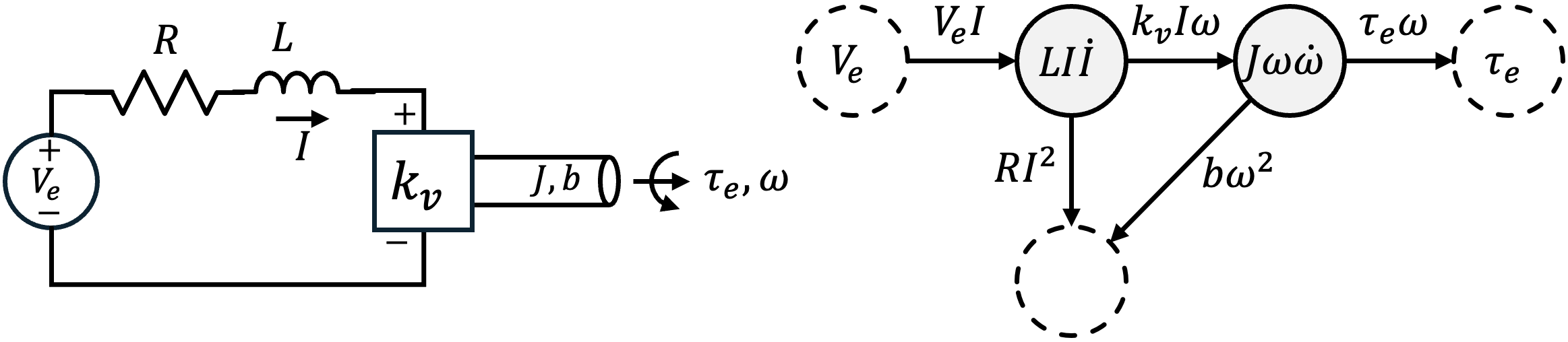}
        \caption{Electric motor}
        \label{fig:sub3}
    \end{subfigure}
    \caption{Example graph-based models for (a) a mass-spring-damper, (b) a buck-converter, and (c) an electric motor.}
    \label{fig:EMgraphexamples}
    \vspace*{-\baselineskip}
\end{figure}

These three common power flow representations can be used to model a wide variety of mechanical and electrical devices. Fig.~\ref{fig:EMgraphexamples} illustrates some common mechanical and electrical systems and components.

\section{Applications to Control and Design Optimization} \label{sec:applications}

Beyond offering a convenient way to construct the analytical governing equations for a multi-domain energy system, the network structure encoded in graph-based models (e.g., via the capacitance matrix $\mat{C}$ and incidence matrix $\widebar{\mat{M}}$ in~\eqref{eq:GBM dynamics-DPin}) can be directly exploited for analysis, control design, estimation, design optimization, and other purposes. This section highlights several examples from the recent literature, spanning both theoretical and applied contributions. Section~\ref{sec:control} focuses on feedback control, while Section~\ref{sec:design} focuses on design optimization.

\subsection{Decentralized and Hierarchical MPC} \label{sec:control}

Graph-based models provide a structure that can be used to develop and analyze feedback control strategies. Graph-based modeling has been primarily developed to capture the dynamics of multi-domain and multi-timescale energy systems, where the control objectives balance performance, efficiency, and safety. The operation of energy systems is often limited by both input constraints as well as key state and output constraints that must be satisfied to maintain safety. However, maximizing performance and efficiency often requires operating the system close to these safety constraints, where conservatism in the control design to guarantee constraint satisfaction can unnecessarily reduce performance and efficiency. 

Therefore, centralized, decentralized, and hierarchical control approaches based on model predictive control (MPC) have been developed that maximize performance and efficiency while explicitly accounting for safety constraints. Moreover, the design and closed-loop analysis of these MPC controllers leverage the unique graph structure. The following provides an example of how the graph-based structure helps to establish the passivity of a general class of energy systems, which is then used to guarantee closed-loop stability of decentralized MPC controllers. Then several simulation-based and experimental applications of hierarchical MPC highlight how the graph structure is used to identify the control architecture and reduced-order models used at each level of the control hierarchy.

As shown in \eqref{eq:GBM dynamics-DPin}, the graph-based model captures the structure of a dynamic system in the form of the partitioned incidence matrix $ \widebar{\mat{M}} \in \mathbb{R}^{N_{\mc{V}_{d}} \times N_{\mc{E}}} $ and the vector of power flows $\vec{P} \in \mathbb{R}^{N_{\mc{E}}} $, where each element of $ \vec{P} $ corresponds to the power flow along an edge of the graph. The first theoretical result leveraging this structure shows that graph-based models are passive (in the sense of \cite{khalil_nonlinear_2002}) for a specific choice of inputs and outputs under certain mild assumptions on the power flow equations \cite{Kolen_2017Automatica}. Specifically, let the power flow defined in \eqref{eq:power flow} have the form 
\begin{equation} \label{eq:power flow passivity}
\begin{split}
P_{j} =f_j\left(x_{j}^{\mrm{tail}},x_{j}^{\mrm{head}}\right) + g_j\left(x_{j}^{\mrm{tail}},x_{j}^{\mrm{head}}\right) u_{j} \:,
\end{split}
\end{equation}
where $ x_{j}^{\mrm{tail}} $ and $ x_{j}^{\mrm{head}} $ are the state of the tail and head vertices for edge $ e_j \in \mathcal{E} $. Here the power flow is a nonlinear function of these vertex states but an affine function of control input $ u_{j} $. Additionally, it is assumed that $ f_j $ is Lipschitz, twice continuously differentiable, and $ f_j(0,0) = 0 $ while $ g_j $ is continuous, $ g_j(0,0) = 0 $, and the intersection of the zero sets of $ g_j $ is the origin, i.e., $ \bigcap_j \mathcal{N}(g_j\left(x_{j}^{\mrm{tail}},x_{j}^{\mrm{head}}\right)) = \{0\}$. With the additional structure from \eqref{eq:power flow passivity}, \eqref{eq:GBM dynamics-DPin} can be rewritten as 
\begin{equation} \label{eq:GBM dynamics passivity}
\mat{C}\dot{\vec{x}}(t) = -\widebar{\mat{M}}\vec{F}(\vec{x},\vec{x}^t) -\widebar{\mat{M}}\vec{G}(\vec{x},\vec{x}^t) \vec{u} + D \vec{P}^{in}\:,
\end{equation}
where $ \vec{F}(\vec{x},\vec{x}^t), \vec{G}(\vec{x},\vec{x}^t), \vec{u} \in \mathbb{R}^{N_{\mc{E}}} $ are vectors of elements from \eqref{eq:power flow passivity} for each edge of the graph.

The external power flows $\vec{P}^{in}$ are assumed to directly affect a subset of states $ \vec{x}^{in}$. Additionally, the power flows from the system to the sink states are denoted as $ \vec{P}^{out}$.

In \cite{Kolen_2017Automatica}, this particular structure is used to show that the system \eqref{eq:GBM dynamics passivity} is passive from inputs $ \widebar{\vec{u}} $ to outputs $ \widebar{\vec{y}} $ where
\begin{equation}
    \widebar{\vec{u}} = \begin{bmatrix} \vec{P}^{in} \\ \vec{u} \\ -\vec{x}^t \end{bmatrix}, \quad \widebar{\vec{y}} =  \begin{bmatrix} \vec{x}^{in} \\ \vec{y} \\ \vec{P}^{out}\end{bmatrix},
\end{equation}
and $ \vec{y} = - \vec{G}(\vec{x},\vec{x}^t) ( \widebar{\mat{M}}^\top \vec{x} + \underwidebar{\mat{M}}^\top \vec{x}^t )$.

\begin{figure}[t]
    \centering
    \includegraphics[width=0.6\linewidth]{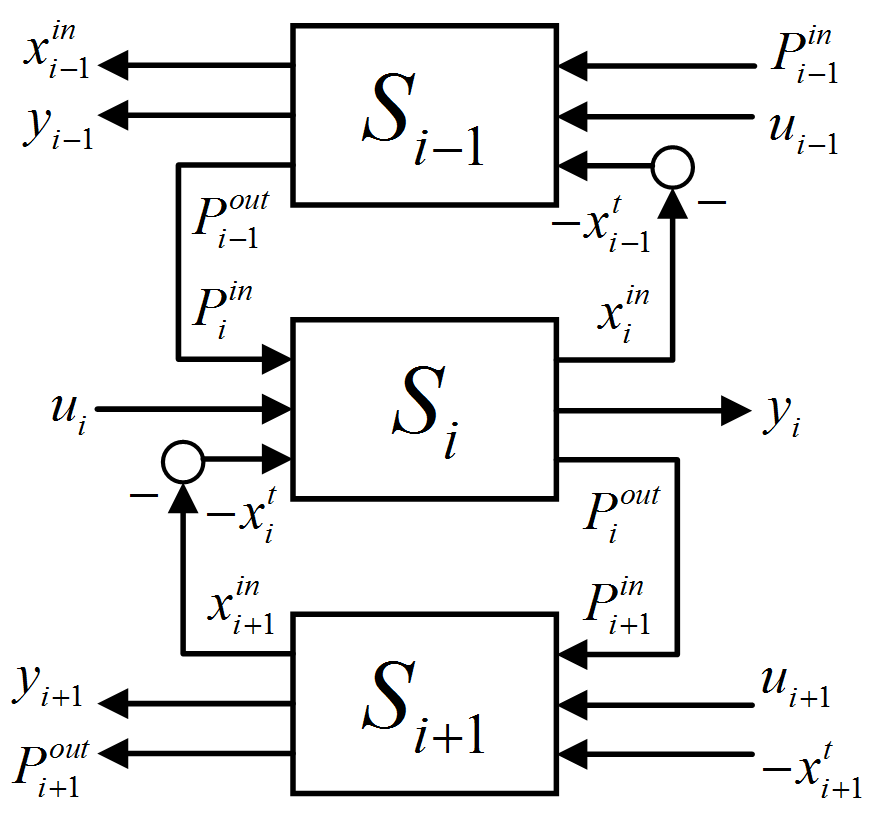}
    \caption{Block diagram showing the negative feedback connection created by power flow between passive subsystems \cite{Kolen_2017Automatica}.}
    \vspace*{-\baselineskip}
    \label{fig:Passivity_Subsystems}
\end{figure}

As shown in Fig. \ref{fig:Passivity_Subsystems}, the passivity of individual subsystems can be established using similar definitions of inputs and outputs, where the dynamic coupling between subsystems creates a negative feedback loop. Thus, a dynamic system can be proven to be passive if each of its subsystems are passive, leveraging the fact that passivity is preserved under negative feedback. 

\begin{figure}[t]
    \centering
    \includegraphics[width=0.8\linewidth]{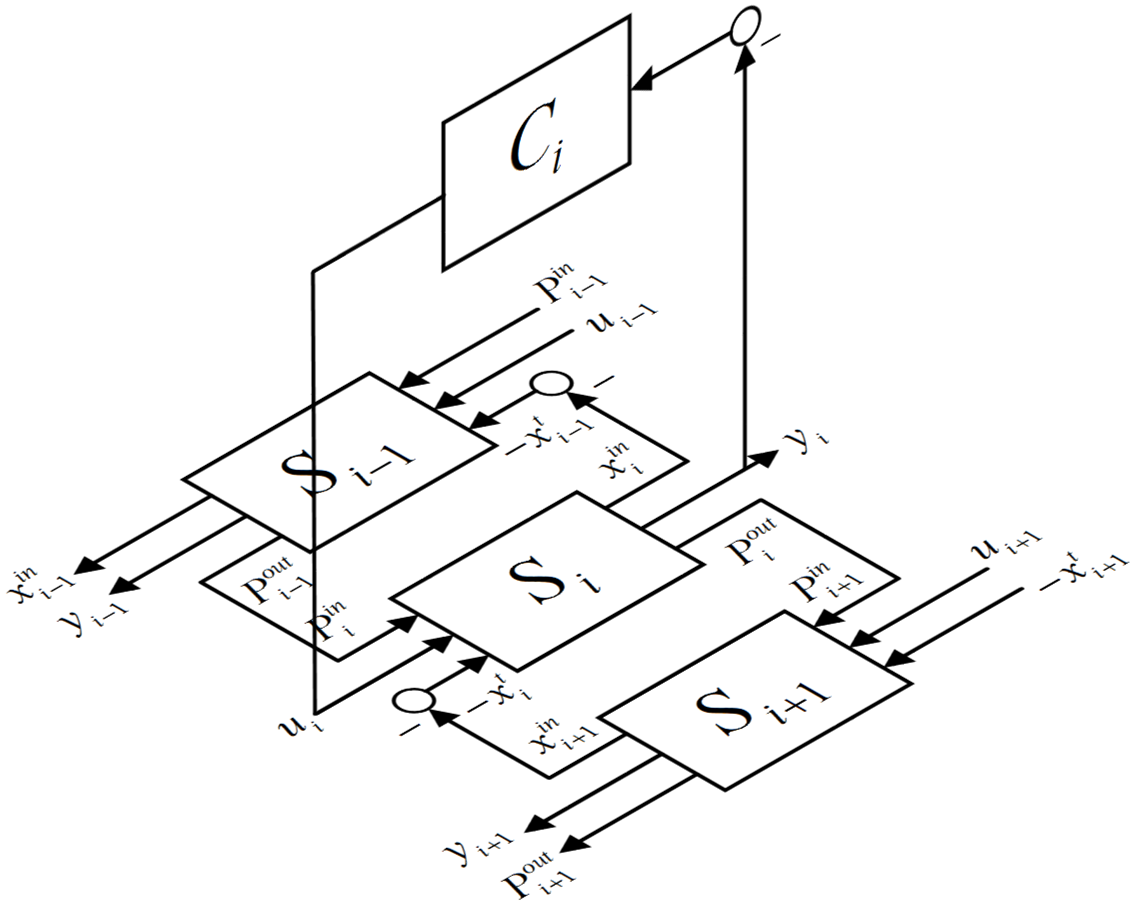}
    \caption{Block diagram showing the negative feedback connection between subsystem $\mathcal{S}_i$ and its MPC controller $ \mathcal{C}_i $ designed with the passivity constraint \eqref{eq:passivity_constraint} \cite{Kolen_2017Automatica}.}
    \vspace*{-\baselineskip}
    \label{fig:Passivity_Control}
\end{figure}

Moreover, the closed-loop system is proven to be passive and, therefore, stable under decentralized control if the controller for each subsystem $\mathcal{S}_i$ is designed to be passive from $\vec{y}_i$ to $ \vec{u}_i$, as shown in Fig. \ref{fig:Passivity_Control}. To ensure the passivity of each subsystem controller, the MPC formulation can be augmented with the constraints
\begin{equation} \label{eq:passivity_constraint}
    \dot{z}_i = \vec{u}_i^\top \vec{y}_i\;, \quad z(\tau) \leq \beta_i \;, \quad \tau \in [0,T]\;,
\end{equation}
where $ z \in \mathbb{R} $ is an auxiliary state representing the accumulation of passivity and $ \beta_i > 0 $ is a predetermined finite constant that limits the time the controller is allowed to operate with a deficiency of passivity. 

In this example, graph-based modeling provided the structure needed to identify the passivity of these systems and developed decentralized MPC controllers where the closed-loop system is guaranteed to be stable. This work has been extended to the decentralized control of hybrid dynamic systems, where the power flow along edges can be switched on and off \cite{Pangborn_2018ACC}. A separate branch of theoretical analysis in \cite{PangbornACC2019,Pangborn2019} established conditions under which switched graph-based models are cooperative (in the sense of~\cite{Angeli2003}) and leveraged this to guarantee robustness of state constraints under a hierarchical MPC framework.

Beyond providing valuable approaches for analysis and theoretical guarantees, the structure of graph-based models also offers practical value in developing hierarchical controllers for multi-timescale dynamic systems. Graph-based models were first used for hierarchical control in \cite{Koeln2015DSCC,Williams2015DSCC}, where \cite{Koeln2015DSCC} proposes hierarchical MPC to specifically address the multi-timescale dynamics common to energy systems and \cite{Williams2015DSCC} develops a three-level hierarchy to control the engine, electrical, and thermal systems of a simulated aircraft. In these initial efforts, the graph structure of the system dynamics is used to manually decompose into multiple dynamically-coupled subsystems. At the lowest level of the hierarchy, the system is decomposed into many small subsystems, each with its own MPC-based controller designed with a relatively fast update rate to account for the fastest timescales of the system and to effectively reject unknown disturbances. The higher levels of the hierarchy are formed by the re-composition of subsystem graphs from the lower levels and order reduction to remove states associated with the fast timescale dynamics. While not directly associated with the eigenvalues of the system, each system state in the graph-based approach has an associated vertex and capacitance. In this work, the magnitudes of these capacitances are used to label each vertex as either fast, medium, or slow. Moving up the three-level hierarchy, fast vertices are treated as algebraic at the middle level of the hierarchy to reduce the number of dynamic states, while both fast and medium vertices are treated as algebraic at the upper-most level to allow the single controller at the top of the hierarchy to focus on planning the behavior of the slowest timescale dynamics over a long prediction horizon. While these initial efforts produced practical hierarchical MPC controllers, the decomposition of the system graph was a suboptimal manual process and the hierarchical controllers lacked theoretical guarantees of stability or constraint satisfaction. 

Going beyond the initial efforts from \cite{Koeln2015DSCC,Williams2015DSCC}, more systematic methods have been developed that leverage the graph structure of the system model to identify how the model should be decomposed into subsystem models used by controllers at each level of a hierarchical controller. In \cite{Docimo_2018CCTA}, hierarchical agglomerative clustering based on a novel energy-based distance metric explicitly uses the graph structure of the model to identify subsystem decompositions that minimize the uncertainty in power flow between neighboring subsystems. In \cite{Yu_2023Aviation}, Koopman operator theory provides a mechanism to decompose the states of a graph-based model by timescale and inform the design of a hierarchical controller. Ref.~\cite{yu2024learningnetworkeddynamicalmodels} additionally combines Koopman operator theory and deep learning with graph-based representations to learn control-oriented dynamics from data.

A fundamental challenge in the model-based control of energy systems is their strong nonlinearities. For MPC, this often requires either solving nonlinear optimization problems online or approximating nonlinearities by a different form, with Taylor series linearization~\cite{Pangborn_2018JDMC} and piece-wise linear models~\cite{PangbornACC2019} as the most common choices. Alternatively,~\cite{Park_2023CCTA} proposes a ``lifting linearization'' approach that augments the structure of a graph-based model with additional vertices whose dynamics are trained from data to approximate nonlinearities, producing a linear model that still obeys the graph-based framework and respects conservation of energy. This demonstrates that the physics-informed graph structure can provide utility in combination with data-based methods, including cases where multiple empirically-derived parameters like heat transfer coefficients can be simultaneously trained from data rather than addressed individually. 

The graph structure is further leveraged in \cite{Docimo2022ACC}, where control co-design is used to simultaneously optimize physical system parameters along with the architecture and design parameters of a hierarchical MPC controller. Both \cite{Docimo_2018CCTA} and \cite{Docimo2022ACC} demonstrate the value of using graph-based models for hierarchical control design when applied to the modeling and control of the power systems for electric vehicles. 

Centralized and hierarchical MPC controllers using graph-based models for prediction have been experimentally validated in several studies, including~\cite{Pangborn2017,Pangborn_2018JDMC,Koeln_202CST} with single-phase thermal management testbeds,~\cite{Aksland_2021CEP} with a UAV powertrain testbed, and~\cite{Aksland_2023CST} with the coupling of a thermal management and UAV powertrain testbed.

\subsection{Dynamic Design Optimization} \label{sec:design}

Graph-based models can be leveraged to optimize energy system sizing and topology while accounting for transient behavior. Requirements for next-generation energy systems, such as increased power production to meet the escalating needs of data centers~\cite{koot2021usage}, have placed a greater emphasis on model-based design tools. Integration of optimization methods with graphical tools is advantageous in that it can provide intuitive correlations between graphs and physical design features of interest. 

The graph-based modeling framework presented in this paper can be used for design optimization as well as control co-design (CCD) to simultaneously define plant and controller features.

\subsubsection{Control Co-Design}
For design and CCD with plant sizing requirements, a four step methodology is recommended to prepare and solve the associated optimization problems~\cite{Docimo_2021JDMC}. Given a baseline model described by (\ref{eq:GBM dynamics}), plant design variables $\vec{\theta}$, and controller design variables $\vec{\phi}$, the first step augments the model into 
\begin{equation} \label{eq:augmented GBM}
\mat{\Psi_c}(\vec{\theta})\mat{C}\dot{\vec{x}}(t) = -\widebar{\mat{M}}\mat{\Psi} (\vec{\theta})\vec{P}(t)\:.
\end{equation}
The diagonal matrix $\mat{\Psi_c}$ serves to scale the size of the vertices and the corresponding storage capacitances. Each entry $i$ of $\mat{\Psi_c}$ contains a function $\psi_{c,i}(\vec{\theta})$, mapping plant design variable values to the dynamics. Similarly, the diagonal matrix $\mat{\Psi}$ serves to scale the edges of the graph, with each entry $j$ of $\mat{\Psi}$ containing a function $\psi_{j}(\vec{\theta})$. These functions and the design variables can be continuous (for sizing) or discrete (for architecture/topology adjustment).

The second and third steps define expressions for the objective and optimization problem constraints, respectively. The objective function $J$ is often a weighted sum of independent or conflicting terms. Example terms include  energy passing through the edges or the integral of state tracking error squared~\cite{Laird_2022Energy}. The most common constraint is the augmented graph-based model of (\ref{eq:augmented GBM}), with defined profiles for $\vec{x^s}(t)$ if performance is to be evaluated under specified environmental conditions. Tied to this is a constraint describing the initial conditions $\vec{x}_0$. If controller design is part of the goals, a control law $\vec{F_c}$ is structured as a function of the vertex states, design variables, references to track, and bounds for states and inputs. Nearly all optimization problems involving graph-based models define lower (\underbar{$\vec{\theta}$}, \underbar{$\vec{\phi}$}) and upper ($\bar{\vec{\theta}}$, $\bar{\vec{\phi}}$) design variable bounds.

The fourth recommended step compiles all of these elements into a full optimization problem. A generalized example with objective function $F_J$ is
\begin{gather}
\label{eq:CCD}
\begin{aligned}
\min_{\vec{\theta},\vec{\phi}}  \quad & J = \int_{0}^{t_f} F_J(\tau,\vec{\theta},\vec{\phi}) \,d\tau \\
\textrm{s.t.:} \quad & \mat{\Psi_c}(\vec{\theta}) \mat{C} \dot{\vec{x}}(t)= -\widebar{\mat{M}} \mat{\Psi} (\vec{\theta}) \vec{P}(t)\\
& \vec{x}(0)=\vec{x}_0\\
& \vec{u}(t)=\vec{F_c}(\vec{x}(t),\vec{x^s}(t),\vec{\theta},\vec{\phi})\\
& \underbar{$\vec{\theta}$} \leq \vec{\theta} \leq \bar{\vec{\theta}}\\
& \underbar{$\vec{\phi}$} \leq \vec{\phi} \leq \bar{\vec{\phi}}\:,
\end{aligned}
\end{gather}
with all time-dependent constraints applied from $t=0$ to final time $t=t_f$. Note that $\vec{x}(t)$, $\vec{u}(t)$, and $\vec{P}(t)$ are implicit decision variables to be determined from the relevant equality constraints and power flow expressions. A variety of solution methods are available to solve (\ref{eq:CCD}), such as combining forward simulation of the dynamics for a given set of design variable values with a genetic algorithm (GA). Solution algorithms can take advantage of the modular nature of the graph-based models to decompose the optimization problem into smaller subproblems~\cite{Smith_2025ACC}. To show how graph-based models can be used for design, a few examples are highlighted below.

\begin{figure}[t]
\centering
\includegraphics[width=1.00\columnwidth]{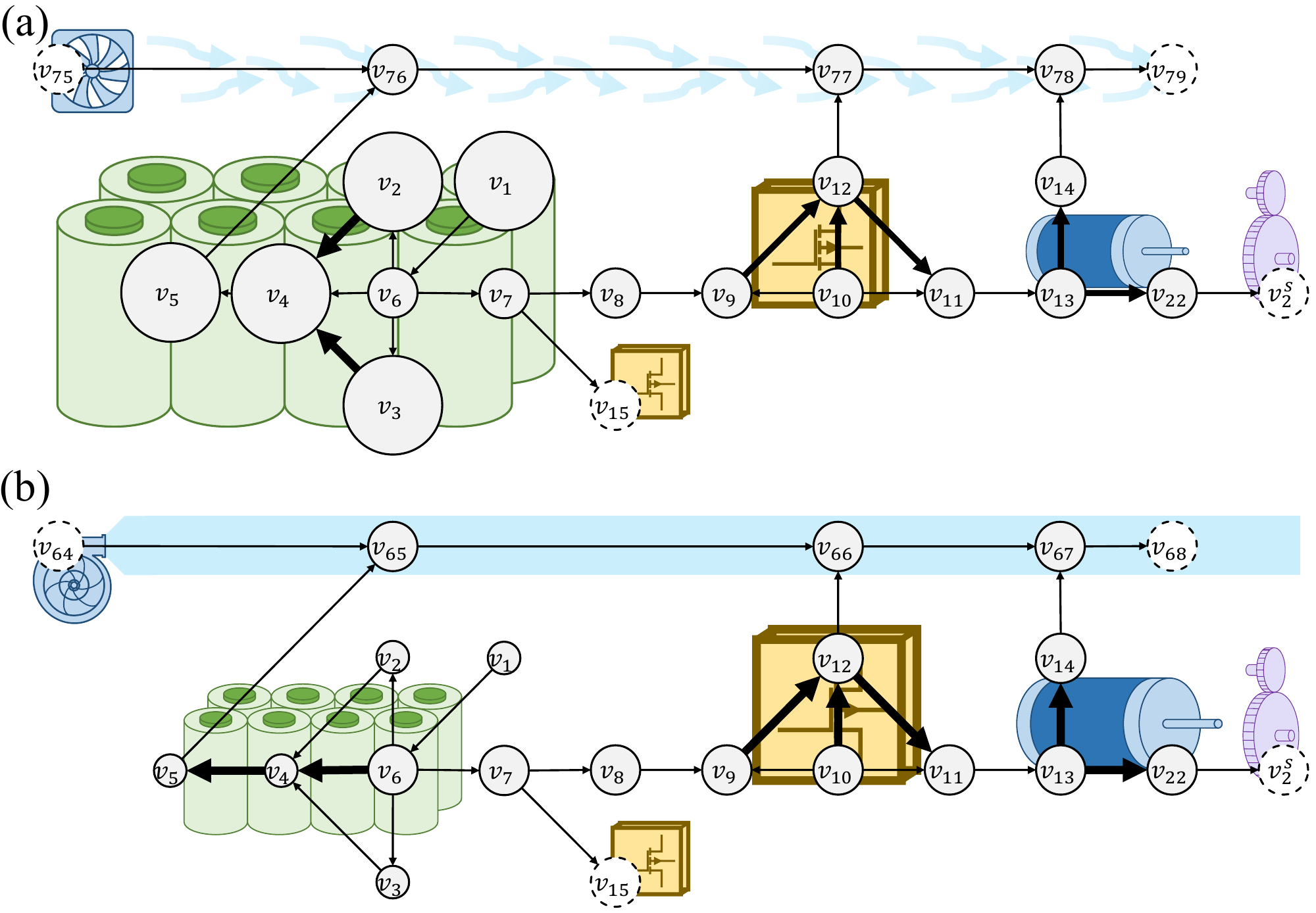}
 \caption{Electric vehicle powertrain subsystem design changes from (a) air cooling to (b) liquid cooling with a reduced battery pack size, achieved by resizing vertices and selecting connecting edges through optimization methods.}
\label{fig:Design Opt}
\end{figure}

The first example focuses on a scenario where only plant parameters and features need to be determined, with the control law $\vec{F_c}$ fixed using predefined trajectories. Fig.~\ref{fig:Design Opt} presents a visual representation of a graph-based model before and after applying the outlined methodology. For this electric vehicle (EV) design problem~\cite{Docimo_2020JMD}, using (\ref{eq:CCD}) leads to a reduced size of the battery pack, as well as selection of air vs. liquid cooling for individual components. 

\begin{figure}[t]
\centering
\includegraphics[width=.9\columnwidth]{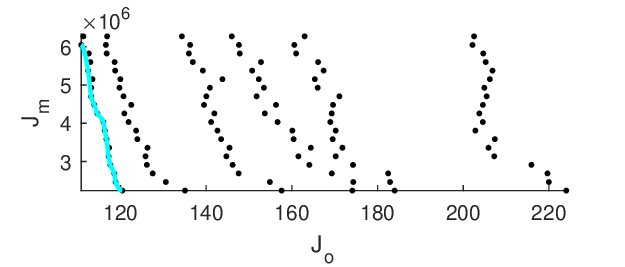}
 \caption{Design options (black) and Pareto front (cyan) comparing manufacturing-stage ($J_m$) and operation-stage ($J_o$) environmental impact in kg of CO\textsubscript{2} equivalent~\cite{jahan2024study}.}
 \vspace*{-\baselineskip}
\label{fig:CCD Results}
\end{figure}

The second example focuses on CCD, showing how the outlined methodology facilitates design for even nontraditional objectives and stationary systems. Microgrids are increasingly developed to solely power data centers. Not only must these provide sufficient and regulated power, but there are expectations that these systems must be manufactured and operated without harmful effect on local and global environments. Integration of environmental sustainability goals for microgrids and data centers is supported by graph-based modeling with CCD. Fig.~\ref{fig:CCD Results} shows how the outcomes of applying the four step methodology informs a tradeoff analysis: the Pareto front contains designs with different levels of balancing between manufacturing-stage greenhouse gas production and waste heat generated during operation. An insight provided by this CCD study is that reducing greenhouse gas production from manufacturing should be emphasized over reducing waste heat from an environmental standpoint~\cite{jahan2024study}.

\subsubsection{Architecture Design Exploration} 
Another aspect of design optimization concerns examining the system architecture itself and how it influences the dynamic behavior of the system. Given a set of components, enumeration algorithms can be used to enumerate all feasible candidate architectures. While it would be extremely time-intensive to rederive the governing equations for each of these candidate system architectures, given graph-based representations of each individual component, ``stitching'' algorithms can be designed to construct a representation of the dynamics of the system using the individual component graphs. This was first explored with the graph-based modeling framework in~\cite{10.1115/1.4043203} for a specific class of single-phase cooling systems, where the candidate graph architectures were enumerated via encoding in labeled rooted trees. In \cite{Vyas_2026SCITECH}, one method to do this is proposed that couples the dynamics of single-state graph models, multi-state graph models, and algebraic models (not represented by graphs). The method is summarized here.

\begin{figure}[t]
    \centering
    \includegraphics[width=1\columnwidth]{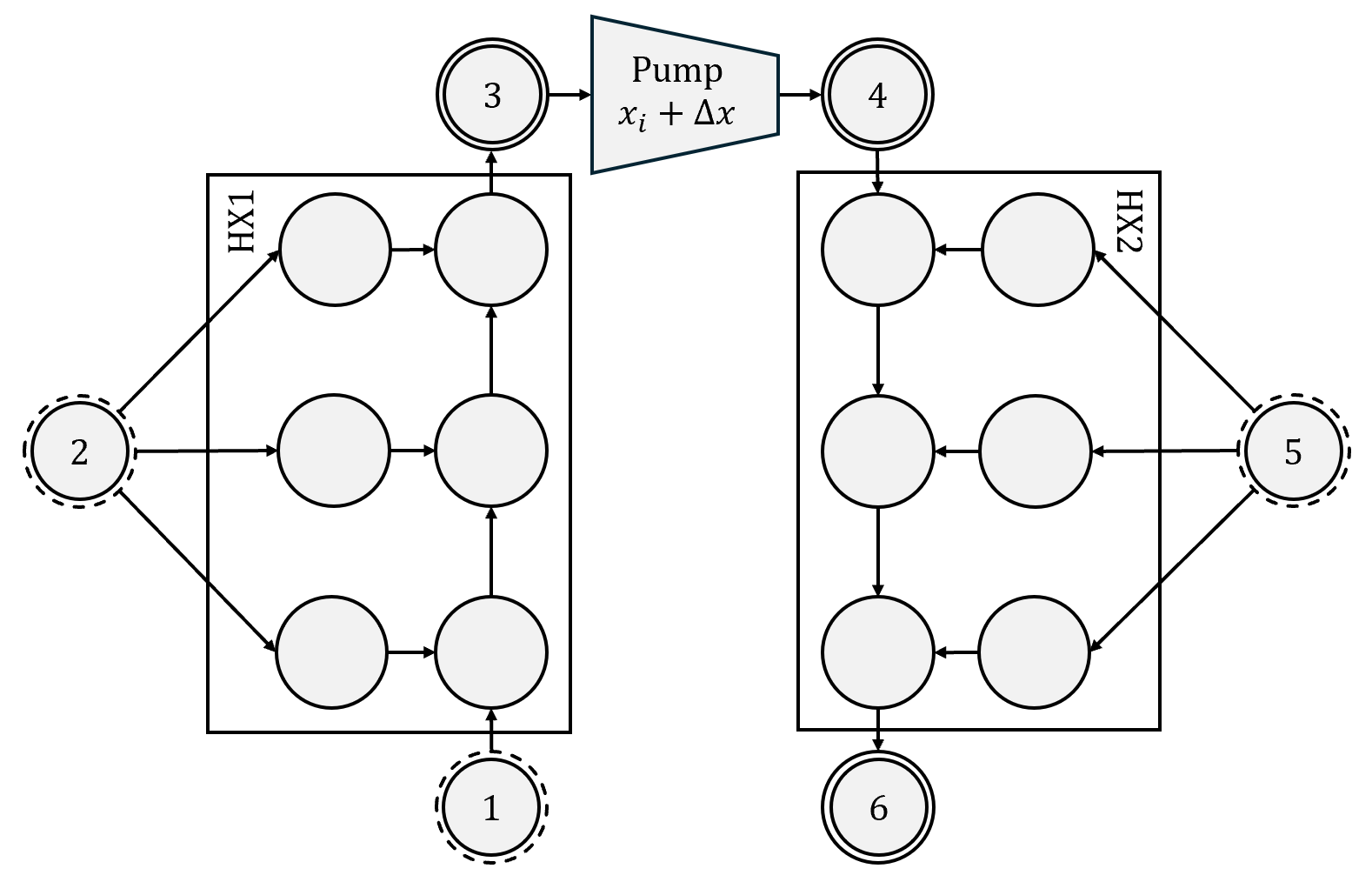}
    \caption{System graph corresponding to formulation with boundary condition (dashed), algebraic (double-circled) and dynamic (solid) vertices.}
    \vspace*{-\baselineskip}
    \label{fig:ExampleSys}
\end{figure}

Given architecture connectivity information, component models are merged into a system graph with algebraic vertices (dashed) and dynamic states (solid) (see Fig.~\ref{fig:ExampleSys}). Algebraic vertices included component graph sink vertices and exogenous boundary condition sink vertices. Algebraic models directly compute outlet states as a function of the inlet states and parameters, but all algebraic vertices  are fundamentally governed by algebraic relationships. The combination of algebraic models and dynamic graph-based models results in a DAE system that is used to accurately capture system transients. In system graphs, algebraic vertices that follow dynamic vertices are assigned an identical state to enforce continuity. The algebraic and dynamic states are concatenated into a global state vector $\vec{x}$, defined as
\begin{equation}
\vec{x}=
\begin{bmatrix}
\vec{x}_{alg}\\\vec{x}_{dyn}
\end{bmatrix}\;,
\end{equation}
where $\vec{x}_{alg}$ and $\vec{x}_{dyn}$ capture algebraic ``states'' and dynamic states respectively.

\begin{figure}[tbp]
    \centering
    \includegraphics[width=.5\columnwidth]{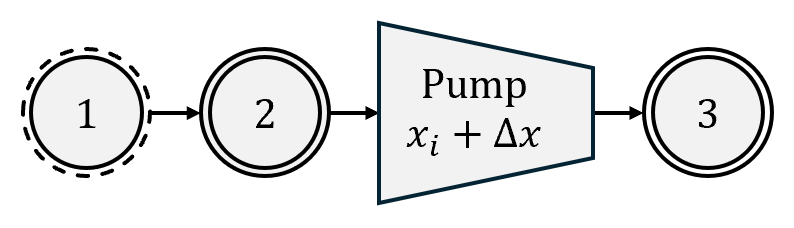}
    \caption{Algebraic model with boundary condition and algebraic vertices.}
    \vspace*{-\baselineskip}
    \label{fig:ExampleF}
\end{figure}

Unlike dynamic graph-based models, algebraic models do not have mathematically defined derivatives. As such, $\dot{\vec{x}}_{dyn}$ contains calculable expressions while $\dot{\vec{x}}_{alg}$ contributes entries of 0, such that
\begin{equation}
\dot {\vec{x}}=
\begin{bmatrix}
\dot {\vec{x}}_{alg}\\\dot {\vec{x}}_{dyn}
\end{bmatrix}=
\begin{bmatrix}
\vec{0}\\\dot {\vec{x}}_{dyn}
\end{bmatrix}\;,
\end{equation}
where 
\begin{equation}
\vec{C}_{dyn}'\dot {\vec{x}}_{dyn}=-(\mat{M}*\mat{S})_{dyn}\vec{\Gamma}_{dyn}\;.
\end{equation}
A vector $\vec{F}$ is defined that characterizes algebraic relationships that govern algebraic states. As such, $\vec{F}_{alg}$ now contains calculable expressions while $\vec{F}_{dyn}$ contributes entries of 0, such that
\begin{equation}
\vec{F}=
\begin{bmatrix}
\vec{F}_{alg}\\\vec{F}_{dyn}
\end{bmatrix}=
\begin{bmatrix}
\vec{F}_{alg}\\\vec{0}
\end{bmatrix}\;.
\end{equation}
For the example shown in Fig.~\ref{fig:ExampleF}, $\vec{F}_{alg}$ is given by
\begin{equation}
\vec{F}_{alg}=
\begin{bmatrix}
x_1-BC\\x_2-x_1\\x_3-(x_2+\Delta x_{pump})
\end{bmatrix}\;,
\label{eq:ExampleF}
\end{equation}
where $BC$ is an exogenous boundary condition model input.
The rows of equation~\eqref{eq:ExampleF} correspond to three types of algebraic constraints: prescription, continuity, and model. A square mass matrix $\mat{S}_{mass}$ is used by a DAE solver internally to distinguish between algebraic and dynamic states. A system with $N_a$ algebraic states and $N_d$ dynamic states defines the mass matrix as
\begin{equation}\mat{S}_{mass}=
\begin{bmatrix}
\textbf{0}_{N_a\times N_a} & \textbf{0}_{N_a\times N_d}\\\textbf{0}_{N_d\times N_a} & \textbf{I}_{N_d}
\end{bmatrix} \;.
\end{equation}
A vector $\vec{R}$ representing the system is given by
\begin{equation}
\vec{R}=\mat{S}_{mass}\dot {\vec{x}}-\vec{F}\;,
\end{equation}
where $\vec{R}$ contains all information necessary for a DAE solver to stitch components and simulate a system graph given component models and an arbitrary system configuration.

\begin{figure}[bt!]
    \centering
    \includegraphics[width=0.5\textwidth]{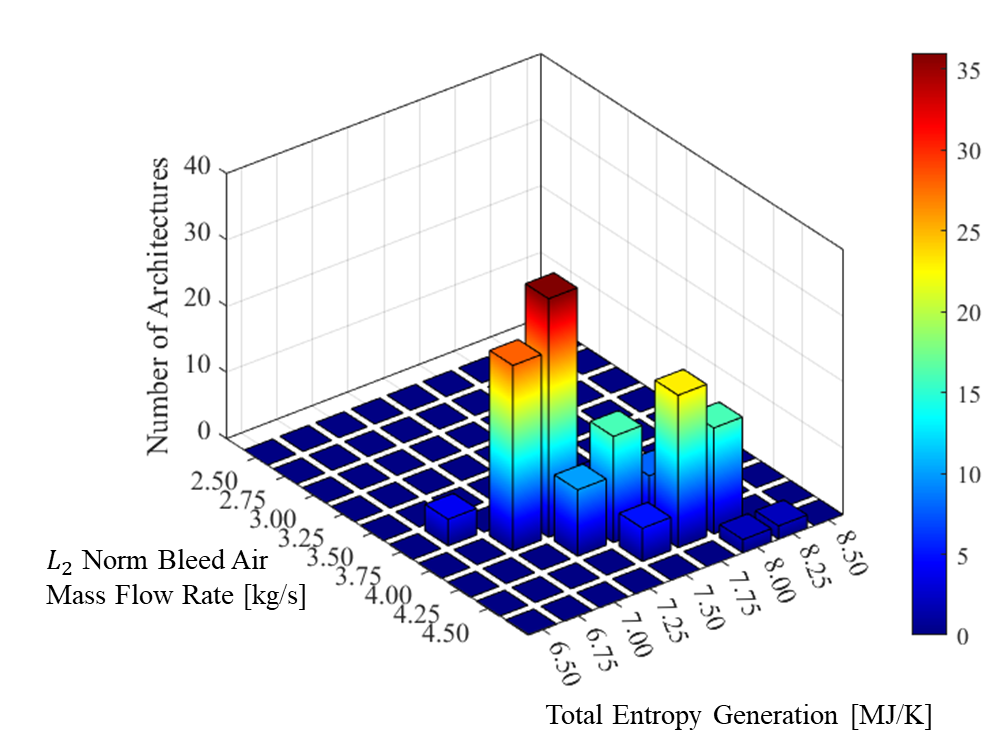}
    \caption{3D histogram depicting the distribution of 164 system architectures, each optimized to minimize total entropy generation over a dynamic mission profile, with respect to  total entropy generated  and $L_2$ norm of bleed air mass flow rate \cite{Bolander_2025TTE}.}
    \vspace*{-\baselineskip}
    \label{fig:entropy_bleedhisto}
\end{figure}

This method enables transient simulation of the system graph, but also can be used for design exploration or optimization over each architecture. This was done for single-phase components in \cite{Bolander_2025TTE} to find the most efficient system architecture, and its associated parameters, with respect to both first and second law efficiency metrics (Fig.~\ref{fig:entropy_bleedhisto}). As in the case of CCD discussed earlier, a feature of this approach is that it enables \textit{dynamic} design optimization of the thermal system.
\section{Toolbox Overview} \label{sec:toolbox}
While the previous sections highlighted some of the concepts and applications of graph based models, many of the results presented in the references came from independently developed and closed-source codebases.  To aid in the adoption of the idea, an open source toolbox would be useful.  The graph-based modeling toolbox (GMT) is a MATLAB-based toolbox designed to streamline the workflow required to leverage graph-based models in control synthesis and optimization methods, as described in previous sections. Users generate system graphs and their associated dynamic equations by connecting elements from a library of component models. The toolbox development has been an aggregation of capabilities and codebases created over the past decade. GMT is built for MATLAB R2025A or later, along with Simulink, Stateflow, and the Symbolic Math Toolbox. GMT is installed using a \verb|.mltbx| file, which installs the toolbox as a MATLAB add-on. The toolbox is available open source and located at \href{https://github.com/psu-PAC-Lab/graph-model-toolbox}{github.com/psu-PAC-Lab/graph-model-toolbox}. 

GMT uses object-orient programming (OOP) to synthesize system and component models from graph elements. Key OOP definitions utilized in this section are classes, superclasses, subclasses, objects, properties, and methods. Classes are general structures composed of properties and methods. Superclasses and subclasses describe the hierarchical relationship between classes. The key concept is that a subclass inherits the superclass methods and properties. Methods are functions that operate on a class and properties are data associated with a class. Lastly, an object is a unique instance of a class. 

This section present examples from the initial open-source release to demonstrate using toolbox classes, methods, and features. Users are expected to reference the latest toolbox documentation descriptions, as future releases may change descriptions. For readability, workspace variables are assigned for each input to a given class method and subsequently passed to the respective method inputs. Object instances are created using constructor methods, which share the same name as their respective classes.

\subsection{Main Toolbox Classes}\label{gbm_classes}
There are several classes within the toolbox to facilitate modeling. The main toolbox classes are arranged in a hierarchical structure to represent their relationships. Table \ref{tab:gmt_mainclass} lists each toolbox class and its description. Both component and system models are composed of the same classes. Most classes in the toolbox have optional input arguments associated with each class and follow the same structure when exercising. When using optional input arguments in the toolbox, each option requires two arguments: The first specifies the option being used and is entered as a string data type, while the second specifies the data associated with that option, which has a unique data type. All optional inputs must be entered after the required class inputs have been entered. Multiple optional input arguments can be used concurrently by concatenating them.  

\begin{table}[tb]
    \centering
    \caption{Main toolbox classes.}
    \label{tab:gmt_mainclass}
    \begin{tabular}{cc}
        \toprule
        \textbf{Class} & \textbf{Description}\\  
        \midrule
        \verb|gmt_Graph| & Graph Model \\ 
        \verb|gmt_Edge| & Graph Edge \\
        \verb|gmt_vertices| & Graph Vertex \\
        \verb|gmt_Parameter| & Model Parameter \\
        \verb|gmt_Input| & Model Input \\
        \verb|gmt_Port| & Connection Port\\
        \bottomrule
    \end{tabular}
\end{table}

\subsection{Generating Object from Component Library}\label{gmt_component}
Each component model in the library is its own subclass of the \verb|gmt_Graph| superclass. Subclasses inherit the properties and methods from the superclass. Most component models only require a string data type input argument to describe the object being instantiated. Some components require additional input arguments, which can be found in the help section of the toolbox. Objects are created by calling the respective component class constructor method with its required input arguments. The constructor method output is the instantiated object, which must be assigned a workspace variable name to be used in the future. A recommend practice is to match the constructor method name input argument with the workspace variable name to aid in connecting component models. A thermal-fluid tank object is created below from the \verb|gmt_Tank| class. The variable \verb|name| is assigned to \verb|"mainTank"| and provided as an input argument to \verb|gmt_Tank| constructor. The resulting object is stored in the workspace variable \verb|mainTank|.
\begin{lstlisting}
>> name = "mainTank" 
>> mainTank = gmt_Tank(name)
\end{lstlisting}

\subsection{Built-In Graph-Based Model Reports and Visualization}
The generalized framework that OOP provides allows the toolbox to standardize report generation and visualization within the toolbox by calling the associated method from any object that is a \verb|gmt_Graph| class or subclass. Both component and system models use the same method and syntax to execute. Table \ref{tab:gmt_reports} outlines the built-in graph-based model reports and visualization. A full report for the \verb|MainTank| object created in Section \ref{gmt_component} can be generated by calling the \verb|gmt_ReportFull| method from the \verb|MainTank| object.

\begin{lstlisting}
>> MainTank.gmt_ReportFull
\end{lstlisting}

\begin{table}[tb!]
    \centering
    \caption{Built-in reports and visualization.}
    \label{tab:gmt_reports}
    \begin{tabular}{cc}
        \toprule
        \textbf{Method} & \textbf{Function}\\  
        \midrule
        \verb|gmt_ReportGraph| & Tabular Edge and Vertex Data \\ 
        \verb|gmt_ReportParameter| & Tabular Model Parameter Data \\
        \verb|gmt_ReportInput| & Tabular Model Input Data \\
        \verb|gmt_ReportPort| & Tabular Connection Port Data \\
        \verb|gmt_ReportInitCon| & Tabular Initial Condition Data \\
        \verb|gmt_ReportFull| & Generates All Reports \\
        \verb|gmt_PlotGraph| & Plots and Labels Graph Figure \\
        \bottomrule
    \end{tabular}
    \vspace*{-\baselineskip}
\end{table}

\subsection{Connecting Component Models}
A static method \verb|gmt_Combine| from the \verb|gmt_Graph| class is used to connect multiple component models. A static method does not require an object when calling but must reference the associated class. The \verb|gmt_Combine| method has three input arguments. The first input is a string data type representing the name of system. The second input is an $n \times 2$ cell array in which each cell contains a component object defined in the workspace. The objects in the first column have priority during the connection process, meaning all properties associated with the connected edge or vertex will be inherited from the primary object. The third input is an $n \times 2$ numerical array specifying the port numbers, where each port corresponds to a component class edge or vertex number associated with each object in the component cell array. The method returns the resulting system object.

Tank and heat load component models are connected together below by connecting the tank output advection on port two to the heat load input advection on port one. A new thermal-fluid heat load object is created from the \verb|gmt_HeatLoad| class. The variable \verb|name| is assigned to \verb|"heatLoad"| and provided as an input argument to \verb|gmt_HeatLoad| constructor. The resulting object is stored in the workspace variable \verb|heatLoad|. The variable \verb|nameSys| is assigned to \verb|"Sys"|. The  variable \verb|Cmp| is assigned to a cell array where the first column is the \verb|mainTank| object, created in Section \ref{gmt_component}, and the second column is the \verb|heatLoad| object. The variable \verb|Prt| defines a numerical array where the first column contains the number $2$ corresponding to port two of \verb|mainTank| object and the second column contains the number $1$ corresponding to port one of \verb|heatLoad| object. Variables \verb|nameSys|, \verb|Cmp|, \verb|Prt| are passed as input arguments to the \verb|gmt_Combine| static method from the \verb|gmt_Graph| class. The returned system object is assigned to workspace variable \verb|Sys|. 

\begin{lstlisting}
>> name = "HeatLoad"
>> heatLoad = gmt_HeatLoad(name)
>> nameSys = "Sys"
>> Cmp = {{mainTank},{heatLoad}}
>> Prt = [2, 1]
>> Sys = gmt_Graph.gmt_Combine(nameSys,Cmp,Prt)
\end{lstlisting}

\subsection{Defining Initial Conditions}
Initial conditions can be defined during or after object instantiation. An initial condition report aids in understanding the associated vertex number, description, units, and if assigned, the initial condition value. All initial conditions must be specified when assigned. Initials conditions can be added when instantiating a component object using the optional input argument \verb|"InitCon"| followed by the initial condition vector. Another method adds initial conditions after a component is assigned to the workspace. Initial conditions are added to the thermal-fluid tank object \verb|MainTank| below after it was instantiated in Section~\ref{gmt_component}. The variable \verb|ic_val| is assigned to a row vector with 300 in the first column and 600 in the second column, corresponding to the first and second state, respectively. The variable \verb|ic_cal| is passed as an input argument to the \verb|gmt_InitCond| method called from the \verb|mainTank| object in the workspace. The method adds the initial conditions and the output is assigned to \verb|mainTank|. 

\begin{lstlisting}
>> ic_val = [300,6000]
>> mainTank = mainTank.gmt_InitCon(ic_val)
\end{lstlisting}

\subsection{Creating a New Component Model}
Component models are subclasses from the \verb|gmt_Graph| superclass that consist of an edge, vertex, input, parameter, port, and edge to vertex pair arrays. All descriptions defined in edges, vertices, parameters, and inputs are used to describe the physical the meaning of the associated object and do not change the underlying mathematics of the model. However, mislabeling descriptions will make the model more difficult to understand, especially in complex systems with many states and power flows. In addition, all equations entered for each class must use standard MATLAB syntax for mathematical expressions.

{Each new component model must define a new class that references the \verb|gmt_Graph| superclass. A template called \verb|gmt_Component| can be found in the component library, which outlines all the required inputs and associated syntax. GMT parses user equations entered as string data types to create the model system of equations. Users create edge and vertex objects from the respective edge and vertex classes. All edge and vertex objects are assumed to be internal to the system unless an optional input argument is entered. 
Externality for edges and vertices can be defined by designating the object as external using the optional input argument, \verb|"External"|, followed by a true logical statement. 

Vertex, input, and parameter classes can have units assigned to each object instance by using the optional input argument, \verb|"Units"|, followed by the string data type designating the units. Similar to descriptions, units are only used to describe the physical meaning of associated object. Readers should refer to Section \ref{gbm_classes} to understand the full requirements for utilizing optional input arguments.} 

An edge class \verb|gmt_Edge| defines the edge equation and describes the edge power flow. The edge class requires two input arguments, with an optional input for externality. The first input argument is a string data type that describes the edge while the second input argument is a string data type that defines the edge equation in terms of the $x^{head}$ and $x^{tail}$ states using the syntax \verb|xh| and \verb|xt|, respectively. The method returns an edge object.

A edge object representing input advection for a thermal-fluid component is created below from the \verb|gmt_Edge| class. The variable \verb|name| is assigned to \verb|"Advection In"|. The variable \verb|edgeEq| is assigned to \verb|"cp_f*u1*xt"| where \verb|cp_f| is the fluid specific heat, \verb|u1| is the mass flow of the edge, and \verb|xt| is the tail state variable connected to the edge. Both variable \verb|name| and \verb|edgeEq| are passed to the \verb|gmt_Edge| constructor method as input arguments. The returned edge object is assigned to the first index of the workspace variable \verb|E|, which represents all the edge objects contained within a given component model class.

\begin{lstlisting}
>> name = "Advection In"
>> edgeEq = "cp_f*u1*xt"
>> E(1) = gmt_Edge(name,edgeEq)
\end{lstlisting}

A vertex class \verb|gmt_Vertex| defines the left-hand side of~\eqref{eq:GBM dynamics conservation law} for each vertex. For dynamic vertices, this is a function of the state derivative $\dot{x}$ and may also be a function of the state $x$ itself. The state derivative can also be omitted from this equation to create an algebraic state. The vertex class requires two input arguments, with optional inputs for externality and units. The first required input argument is a string data type that describes the vertex, while the second input argument is a string data type that defines the vertex equation in terms of the state derivative and the state using the syntax \verb|x_dot| and \verb|x|, respectively. Vertex equations specified using the state derivative in the form \verb|C*x_dot| or \verb|C*x*x_dot|, where \verb|C| is a coefficient representing all or part of the vertex capacitance, will define a dynamic state type, meaning the power flows define a differential equation for the state. In the example \verb|C*x*x_dot|, where both the state and state derivative are specified, the capacitance is defined as \verb|C*x|.

A vertex object representing the fluid temperature for a thermal-fluid component is created below from the \verb|gmt_Vertex| class. The variable \verb|name| is assigned to \verb|"Fluid Temperature"|. The variable \verb|vertexEq| is assigned to \verb|"cp_f*V_f*rho*x_dot"| where \verb|cp_f| is the fluid specific heat parameter, \verb|V_f| is the vertex volume parameter, \verb|rho| is the fluid density parameter, and \verb|x_dot| is the state derivative. The variable \verb|opt1| is assigned to "Units", and the variable \verb|units| is assigned to \verb|"K"|, representing the temperature units in Kelvin. All four variables are passed to the \verb|gmt_Vertex| constructor method as input arguments. The returned vertex object is assigned to the first index of the workspace variable \verb|V|, which represents all the vertex objects contained within a given component model class.

\begin{lstlisting}
>> name = "Fluid Temperature"
>> vertexEq = "cp_f*V_f*rho*x_dot"
>> opt1 = "Units"
>> units = "K"
>> V(1) = gmt_Vertex(name,capEq,opt1,units)
\end{lstlisting}

A parameter class \verb|gmt_Parameter| defines the parameterization variables for a given model and describes the parameter variable within the model. Both scalar and MATLAB native lookup tables are supported. Future support for MATLAB native neural networks is planned, as neural network-based estimation of fluid properties can provide improved balance between computational efficiency and accuracy when compared to lookup tables and polynomial methods \cite{kolen2025ACC}. The class requires three input arguments, with two optional arguments, which define the parameter units and whether a variable is an optimization parameter. The first input argument is a descriptive name for the parameter, the second input argument is a variable name or an expression that relates a variable to a lookup table in terms of the state variables, and the third input is the numeric data associated with the parameter. Both the first and second input arguments are entered as string data types. Scalar parameters assign a scalar numeric data type to the data input whereas lookup tables assign the data in the form of a structure data type. Each field name under the structure must match the variable defined in the parameter variable expression. Parameters defined as optimization variables will not have numeric values substituted in for variables when a final symbolic expression is generated. Section \ref{gbm_export} in the paper outlines the process for obtaining symbolic expressions for models.

{A fluid specific heat parameter object is instantiated below from the \verb|gmt_Parameter| class. Each variable is assigned its respective value: \verb|name| is assigned to \verb|"Fluid Specific Heat"|, \verb|var| is assigned to \verb|cp_f|, \verb|val| is assigned to 3300, \verb|opt1| is assigned to \verb|"Units"| and \verb|units| is assigned to \verb|"kJ/(kg*K)"|. All five variables are passed to the \verb|gmt_Parameter| constructor method as input arguments. The returned parameter object is assigned to the first index of the workspace variable \verb|Pa|, which represents all the parameter objects contained within a given component model class. 
 
\begin{lstlisting}
>> name = "Fluid Specific Heat"
>> var = "cp_f"
>> val = 3300
>> opt1 = "Units"
>> units = "kJ/(kg*K)"
>> Pa(1) = gmt_Parameter(name,var,val,opt1,units)
\end{lstlisting}}

An input class defines the model input variables and describes the input variables within the model. An input class has two required input arguments to create an object, with an optional input to define the units. The first input argument is a string data type that describes the physical meaning of the input, and the second input is a string data type that defines the input variable. Input variable syntax uses the character \verb|u| followed by a number. The best practice is to
number the inputs in numerical order starting from one for each component class.

An input object is created below from the \verb|gmt_Input| class. Each variable
is assigned its respective value: \verb|name| is assigned to \verb|"Inlet Mass Flow 1"|, \verb|var| is assigned to \verb|"u1"|, \verb|opt1| is assigned to \verb|"Units"| and \verb|units| is assigned to \verb|"kg/s"|. All four variables are passed to the \verb|gmt_Input| constructor method as input arguments. The returned input object is assigned to the first index of the workspace variable \verb|I|, which represents all the inputs objects contained within a given component model class. 

\begin{lstlisting}
>> name = "Inlet Mass Flow 1"
>> var = "u1"
>> opt1 = "Units"
>> units = "kg/s"
>> I(1) = gmt_Input(name,var,opt1,units)
\end{lstlisting}

{A port class defines the model connection points, which can be either an edge or a vertex. The port class requires three input arguments to create an object. The first input is a string data type that defines the connection type, which is either an \verb|EdgeConnection| or \verb|VertexConnection|, the second input is a numerical data type that defines the edge or vertex number associated with the connection, and the third input defines the connection energy domain.}

A port object is created below from the \verb|gmt_Port| class. Each variable is assigned its respective value:  \verb|conType| is assigned to \verb|"EdgeConnection"|, \verb|num| is assigned to 1 and \verb|domain| is assigned to \verb|"Thermal"|.  All three variables are passed to the \verb|gmt_Port| constructor method as input arguments. The returned input object is assigned to the first index of the workspace variable \verb|Po|, which represents all the ports objects contained within a given component model class. 

\begin{lstlisting}
>> conType = "EdgeConnection"
>> num = 1
>> domain = "Thermal"
>> Po(1) = gmt_Port(conType,num,domain)
\end{lstlisting}

{An edge matrix defines the edge to vertex relationship for a given model and is used to create the incidence matrix for the model. The edge matrix is a two-column array where the first column is the tail vertex and the second column is the head vertex. An edge matrix variable is created below. The first row in the array defines the edge 1 connection as starting from vertex 2 and ending to vertex 1. The starting vertex corresponds to the tail vertex and the ending vertex corresponds to the head vertex.  

\begin{lstlisting}
>> EdgeMatrix = [2 1];
\end{lstlisting}}

\subsection{Input Dependencies}
GMT does not automatically handle input dependencies because the components within a system could change the input dependencies and a case-specific automation algorithm is expected to be complex. An example would be a thermal-fluid junction with $n$ input mass flow rates and $m$ output mass flow rates. In this case, the output mass flow rates could be a function of the input mass flow rates, a function of a downstream connected component like a pump, or a combination of the two.  Therefore, a simplified approach was selected by creating a method for the user to specify the input dependencies within the model. The input dependency method \verb|gmt_InputCommon| can be called from any \verb|gmt_Graph| object and requires one input argument. The input argument is a two-column string array where the first column is the current input variable and the second column is the new input variable. The new input variable can be a mathematical expression. The inputs for a split junction object called \verb|Split| are combined below. The variable \verb|common| is assigned to a string array \verb|["u3", "(u1+u2)"]|. The object \verb|Split| calls the \verb|gmt_InputCommon| method and passes the variable \verb|common| as an input argument in the method. The returned object is a new object \verb|Split| that replaces every occurrence of \verb|u3| in the expression with \verb|(u1+u2)|. 

\begin{lstlisting}
>> common = ["u3", "(u1+u2)"]
>> Split = Split.gmt_InputCommon(common)
\end{lstlisting}

\subsection{MATLAB Command Line Simulation}
The command line simulation feature allows systems to be simulated directly in the command line. Similar to initial conditions, MATLAB command line simulations can be built during or after object instantiation. The command line simulation feature builds an ODE function and simulation script m-file for the associated object and then saves the files to a user-specified file path. The simulation files for the \verb|mainTank| object created in Section \ref{gmt_component} are built below after the object is instantiated to the workspace. A variable \verb|fp| is assigned to \verb|string(pwd)|, which is the current working directory in MATLAB. The \verb|gmt_BuildSim| method is called from the \verb|mainTank| object with \verb|fp| assigned as an input argument. The expression is assigned to the \verb|mainTank| object to overwrite the existing object in the workspace. The resulting simulation files are stored in a new folder in the current working directory. 

\begin{lstlisting}
>> fp = string(pwd)
>> mainTank = mainTank.gmt_BuildSim(fp)          
\end{lstlisting}

Alternatively, the optional input argument \verb|"BuildSim"| followed by the path name as a second input argument can be used to build the simulation files during object instantiation. 

\subsection{Simulink Simulation}
\begin{figure}[tb!]
    \centering
    \includegraphics[width=.95\columnwidth]{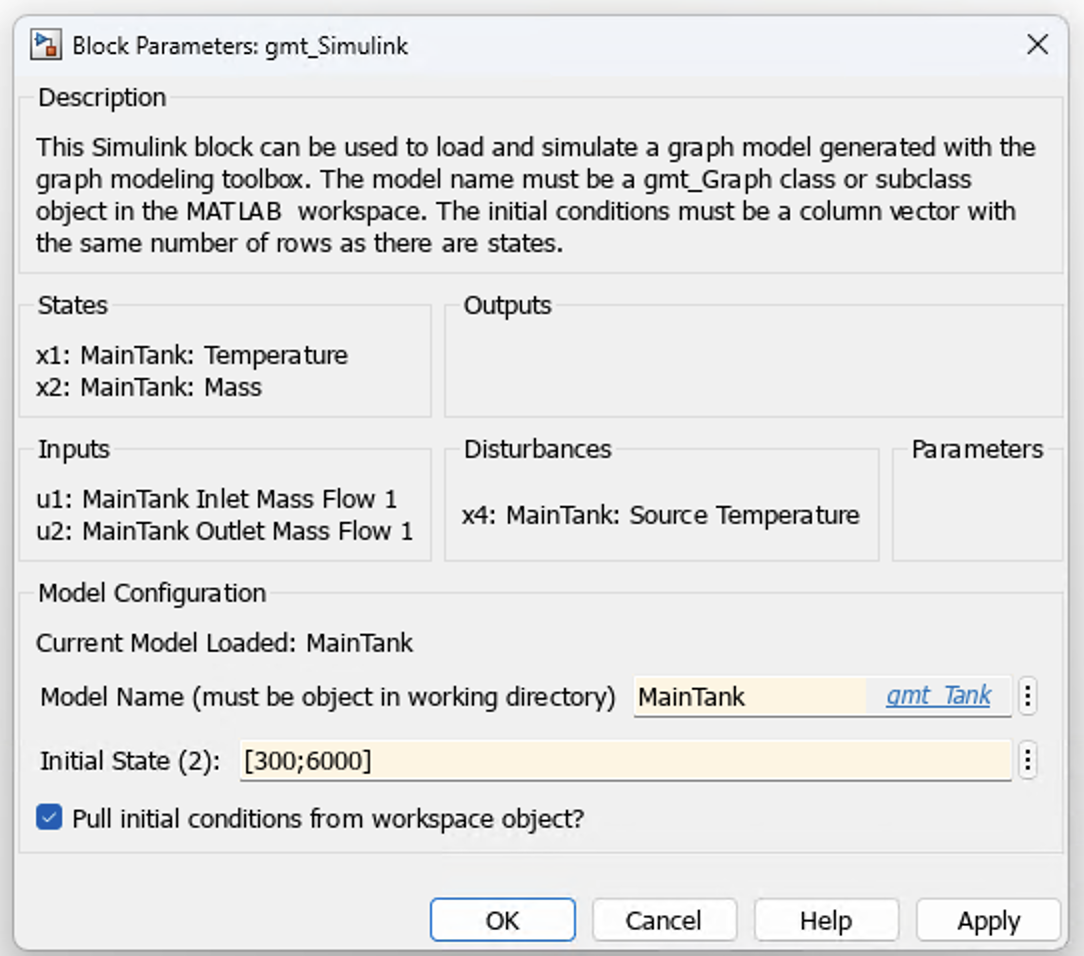}
    \caption{Example Simulink library block mask.}
    \vspace*{-\baselineskip}
    \label{fig:gmt_SimulinkMask}
\end{figure}

A Simulink library block was created to simulate graph-based models in a Simulink environment by computing and integrating the state derivatives as a function of control inputs, disturbances, and user-defined external model parameters. A block mask is used to update the mask variable descriptions, MATLAB function block code, and block port configuration. Users specify the workspace object variable name, initial conditions, and initial condition source in the mask and then the block is updated using the specified workspace object, which must be in the MATLAB workspace and a \verb|gmt_Graph| class or subclass. Users can specify unique initial conditions or use initial conditions in the workspace object by selecting the optional check box. All MATLAB function code originates from the mask defined workspace variable, including any parameters that are not designated as optimization variables. Fig. \ref{fig:gmt_SimulinkMask} shows the updated Simulink mask for a \verb|mainTank| object with initial conditions defined in the workspace object. The model has two control inputs representing the tank inlet and outlet mass flow rates, one disturbance representing the input advection source temperature, no external model parameters, and two states representing the tank mass and fluid temperature.

\subsection{Built-In Control Synthesis}\label{gbm_export}
A method, \verb|gmt_ControlModel|, was developed to provide linear approximation of a nonlinear dynamic system at a point $\vec{x_0}$ and $\vec{u_0}$ using a first-order Taylor expansion. A general nonlinear dynamic system is defined as 
\begin{align}
\vec{\dot{x}} = f(\vec{x},\vec{u}) \;.
\end{align}
After a first-order Taylor expansion, the linear approximation of the nonlinear dynamic system can be expressed as 
\begin{align}
\ \vec{\dot{x}}_{lin} = \mat{A}\vec{x}_{lin}+ \vec{B}\vec{u} + \vec{Z} \;, \label{eq:gmt_linear} 
\end{align}
where $\vec{\dot{x}}_{lin}$ is a column vector of linearized state derivatives expressed in terms of matrices $\mat{A}$ and $\vec{B}$, vector $\vec{Z}$, linearized states $\vec{x}_{lin}$, and control inputs $\vec{u}$. Matrices $\mat{A}$ and $\vec{B}$ are partial derivatives of the nonlinear system dynamics taken with respect to states $\vec{x}$ and control inputs $\vec{u}$, respectively. Both partial derivatives are evaluated at the linearization point $\vec{x}_0$ and $\vec{u}_0$ and are written as 
\begin{align}
\mat{A} = \frac{\partial \vec{f}(\vec{x},\vec{u})}{\partial \vec{x}} \Bigr|_{\substack{\vec{x}=\vec{x}_{0}\\ \vec{u}=\vec{u}_{0}}} \quad
\vec{B} = \frac{\partial \vec{f}(\vec{x},\vec{u})}{\partial \vec{u}} \Bigr|_{\substack{\vec{x}=\vec{x}_{0}\\ \vec{u}=\vec{u}_{0}}} \;.
\end{align}
The vector $\vec{Z}$ is a affine transformation to represent the linear approximation in absolute terms rather than as a deviation from the linearization point. This is defined as 
\begin{align}
\ \vec{Z} = \vec{\dot{x}}_{0} - \mat{A}\vec{x}_{0} - \vec{B}\vec{u}_{0} \;,
\end{align}
where $\vec{\dot{x}_{0}}$ is a column vector of state derivatives computed by evaluating $f(\vec{x}_0,\vec{u}_{0})$. The method returns symbolic expressions for $\mat{A}$, $\mat{B}$, and $\mat{Z}$ as a function of $\vec{x}_0$ and $\vec{u}_0$ to support model-based control design.

\subsection{Equation Exporting}
All \verb|gmt_Graph| class and subclass objects, which represent the model parameterization with analytical expressions, have a system of equations that can be exported for use in other programming languages, such as Python. Two expression options are available for exporting, depending on the requirements. The first is an analytical expression without numerical value substitution for all parameters. The second is an analytical expressions with numerical value substitution for parameters that are not designated as optimization variables. Both expressions are accessed under the \verb|ModelMetadata| property of the object.

\subsection{Future Planned Capability}
GMT has future capability planned to support building a system graph model from a Simulink block diagram of connected components. The feature will be designed to aid engineers in building a graph-based model from a graphical user interface rather than command line. In addition, GMT will have a workflow for creating an MPC formulation, using the graph-based model for prediction.
\section{Conclusion, Future Directions, \\and Open Problems} \label{sec:conclusion}

This paper introduces the use of graph-based models as tools to represent systems that satisfy conservation laws such as conservation of energy or conservation of mass.  While there are certainly approaches that can provide higher-fidelity results than the proposed method, the approach has advantages that would be useful for the control designer to have in their arsenal of tools.  Two of the advantages discussed are (a) the ability to use the tools from graph theory to understand model structure and topology as well as (b) the ability to compartmentalize models in a formal manner.  The second advantage plays well into model-based systems engineering (MBSE) for systems that are under design, allowing for design tradeoffs on various components to be easily made while still in the early design stages of a system.

As illustrated in the text, there are myriad uses of these for a wide range of applications.  Energy domains such as thermal, electrical, mechanical, and chemical have all been previously represented in the literature.  Detailed examples illustrate how the tools are applied to specific components and how these components can be integrated into larger systems.  By providing examples of modeling from both the thermal-fluid domains as well as the electro-mechanical domains, the paper demonstrates the broad applicability of the general approach.

The two advantages listed above are demonstrated to be useful in the control of dynamical systems using various degrees of centralized optimization.  Appropriate decomposition and compartmentalization appropriately determine the right levels of subsystem aggregation for controller design.  This includes the types of multi-timescale dynamics prevalent in energy systems, where subsystems can operate across a wide range of timescales.  Hierarchical control approaches are particularly effective in this case.

In addition to the optimization of the control algorithm, it is possible to use the modularity and composability of systems presented in a graph format to allow for optimization of the plant along with the controller. This is the co-design problem.  Parametric system changes can be represented through edge or vertex parameters, and topological system changes can be represented by graph connectivity.  These can be exploited to aid in the control co-design process more efficiently than many other system model formulations.

Several of these benefits were introduced before in the literature covered by this tutorial.  One key new contribution in this tutorial is the introduction of an open-source toolbox presented to the broader research community to allow easy access to these tools.  The goal is to greatly reduce any barriers to entry for researchers who wish to augment their existing tools with these. Examples from this toolbox are included to help future readers upskill quickly in this domain.

Although significant research has been done with this graph-based modeling approach to date, the field is still quite open for additional contributions.  
 
For example, these tools would benefit from further expansion towards data-driven models beyond the simpler static relationships already incorporated.  From an implementation perspective, greater adoption of the toolbox is highly desired, along with contributions from other researchers that add to the existing component and system models as well as new ways to compose, decompose, and recompose systems efficiently.

In addition to expanding capability and the user base, future goals would include larger-scale case studies on industrial MBSE.  This incorporates the open problem of streamlining the co-design problems for which these tools are well suited.  Co-design demonstrations to date have focused on subsystems; in the future it would be valuable to examine a full system, such as a vehicle, with hundreds or thousands of dynamic states.  At present, the use of existing optimization tools with the graph-based representations encounters a challenge in scaling the state dimension.  In much the same manner that controllers can exploit the structure uncovered through graph analysis, optimization approaches could also be tailored to exploit structure and lead to significantly greater numerical efficiency.  The authors hope that this tutorial paper, along with the available toolbox, will serve as an inflection point in the usage and contribution to the modeling and control engineering community.

\bibliography{references}

@article{1000174,
  author={Fritzson, P. and Bunus, P.},
  journal={Proceedings of the 35th Annual Simulation Symposium}, 
  title={Modelica-a general object-oriented language for continuous and discrete-event system modeling and simulation}, 
  year={2002},
  volume={},
  number={},
  pages={365-380},
  month={Apr.},
  doi={10.1109/SIMSYM.2002.1000174}}

@article{shanks2023design,
  title={Design and validation of a state-dependent Riccati equation filter for state of charge estimation in a latent thermal storage device},
  author={Shanks, Michael and Inyang-Udoh, Uduak and Jain, Neera},
  journal={Journal of Dynamic Systems, Measurement, and Control},
  volume={145},
  number={9},
  pages={091002},
  year={2023},
  publisher={American Society of Mechanical Engineers}
}

@article{Angeli2003,
author = {Angeli, David and Sontag, Eduardo D},
eprint = {0206133v1},
journal = {IEEE Transactions on Automatic Control},
number = {10},
pages = {1684--1698},
primaryClass = {arXiv:math},
title = {{Monotone Control Systems}},
volume = {48},
year = {2003}
}

@ARTICLE{4140745,
  title={Bond-graph modeling}, 
  author={Gawthrop, Peter J. and Bevan, Geraint P.},
  journal={IEEE Control Systems Magazine}, 
  volume={27},
  number={2},
  pages={24-45},
  month={Apr.},
  year={2007},
  doi={10.1109/MCS.2007.338279}
}

@article{SYS-002,
title = {Port-{H}amiltonian Systems Theory: An introductory Overview},
author = {Arjan van der Schaft and Dimitri Jeltsema},
journal = {Foundations and Trends in Systems and Control},
volume = {1},
number = {2-3},
pages = {173-378},
month = {Jun.},
year = {2014},
issn = {2325-6818},
url = {http://dx.doi.org/10.1561/2600000002},
doi = {10.1561/2600000002},
}

@article{Russel_2022IJR,
title = {Graph-Based Dynamic Modeling of Two-Phase Heat Exchangers in Vapor Compression Systems},
author = {Kayla M. Russell and Christopher T. Aksland and Andrew G. Alleyne},
journal = {International Journal of Refrigeration},
volume = {137},
pages = {244-256},
month = {May},
year = {2022},
issn = {0140-7007},
doi = {https://doi.org/10.1016/j.ijrefrig.2022.02.018},
}

@article{10.1115/1.4043203,
    title = {Optimal Flow Control and Single Split Architecture Exploration for Fluid-Based Thermal Management},
    author = {Peddada, Satya R. T. and Herber, Daniel R. and Pangborn, Herschel C. and Alleyne, Andrew G. and Allison, James T.},
    journal = {Journal of Mechanical Design},
    volume = {141},
    number = {8},
    pages = {083401},
    month = {Apr.},
    year = {2019},
    issn = {1050-0472},
    doi = {10.1115/1.4043203},
    url = {https://doi.org/10.1115/1.4043203},
    eprint = {https://asmedigitalcollection.asme.org/mechanicaldesign/article-pdf/141/8/083401/6402048/md_141_8_083401.pdf},
}

@article{Bolander_2025TTE,
  title={A Second Law Approach to Dynamic Optimization of Enumerated Aircraft Thermal Management System Architectures}, 
  author={Bolander, Ara and Bird, Trevor J. and Manion, Reynolds and Glebocki, Marcin and McCarthy, Kevin and Jain, Neera},
  journal={IEEE Transactions on Transportation Electrification}, 
  volume={11},
  number={5},
  pages={12657-12666},
  month = {Oct.},
  year={2025},
  doi={10.1109/TTE.2025.3593116}
  }

@article{Kolen_2017Automatica,
title = {Stability of decentralized model predictive control of graph-based power flow systems via passivity},
author = {Justin P. Koeln and Andrew G. Alleyne},
journal = {Automatica},
volume = {82},
number={},
pages = {29-34},
month = {Aug.},
year = {2017},
doi = {10.1016/j.automatica.2017.04.026},
}

@article{Aksland_2025CSM,
  title={An Approach to Simultaneous Topology, Parametric, and Feedback Control Co-design: Applications to Conservation-Based Energy Systems}, 
  author={Aksland, Christopher T. and Alleyne, Andrew G.},
  journal={IEEE Control Systems Magazine}, 
  volume={45},
  number={3},
  pages={28-55},
  month = {Jun.},
  year={2025},
  doi={10.1109/MCS.2025.3554443}
}

@article{Alleyne2022,
  title     = {Control as an enabler for electrified mobility},
  author    = {Alleyne, Andrew G and Aksland, Christopher T},
  journal   = {Annual Review of Control, Robotics, and Autonomous Systems},
  volume    = {5},
  number    = {},
  pages     = {659--688},
  month     = {May},
  year      = {2022},
  doi       = {10.1146/annurev-control-042920-012513}
}

@article{Docimo_2020JMD,
  title   = {A Novel Framework for Simultaneous Topology and Sizing Optimization of Complex, Multi-Domain Systems-of-Systems},
  author  = {Docimo, Donald J and Kang, Ziliang and James, Kai A and Alleyne, Andrew G},
  journal = {Journal of Mechanical Design},
  volume  = {142},
  number  = {9},
  pages   = {},
  month   = {Sep.},
  year    = {2020},
  doi     = {10.1115/1.4046066}
}

@article{Aksland_2021CEP,
  title   = {Hierarchical model-based predictive controller for a hybrid UAV powertrain},
  author  = {Aksland, Christopher T and Alleyne, Andrew G},
  journal = {Control Engineering Practice},
  volume  = {115},
  number  = {},
  pages   = {104883},
  month   = {Oct.},
  year    = {2021},
  doi     = {10.1016/j.conengprac.2021.104883}
}

@article{Williams_2017JDMC,
  title   = {Dynamical Graph Models of Aircraft Electrical, Thermal, and Turbomachinery Components},
  author  = {Williams, Matthew A and Koeln, Justin P and Pangborn, Herschel C and Alleyne, Andrew G},
  journal = {Journal of Dynamic Systems, Measurement, and Control},
  volume  = {140},
  number  = {4},
  pages   = {},
  month   = {Apr.},
  year    = {2017},
  doi     = {10.1115/1.4038341}
}

@article{Pangborn_2018JDMC,
  title   = {Experimental Validation of Graph-Based Hierarchical Control for Thermal Management},
  author  = {Pangborn, Herschel C and Koeln, Justin P and Williams, Matthew A and Alleyne, Andrew G},
  journal = {Journal of Dynamic Systems, Measurement, and Control},
  volume  = {140},
  number  = {10},
  pages   = {},
  month   = {Oct.},
  year    = {2018},
  doi     = {10.1115/1.4040211}
}

@article{Doty2015b,
    title   = {Dynamic Thermal Management for Aerospace Technology: Review and Outlook},
    author  = {Doty, J and Yerkes, K and Byrd, L and Murthy, J and Alleyne, A and Wolff, M and Heister, S and Fisher, T S},
    journal = {Journal of Thermophysics and Heat Transfer},
    volume  = {31},
    number  = {1},
    pages   = {86--98},
    month   = jan,
    year    = {2017},
    doi     = {10.2514/6.2015-2086},
    isbn = {9781624103438},
    issn = {0887-8722},
}

@article{Kolen_2018Automatica,
  title   ={Robust hierarchical model predictive control of graph-based power flow systems},
  author  ={Koeln, Justin P. and Alleyne, Andrew G.},
  journal ={Automatica},
  volume  ={96},
  number  = {},
  pages   = {127--133},
  month   = oct,
  year    = {2018},
  doi     = {10.1115/1.4064771}
}

@article{Lionello_2020Energy,
  title   = {Graph-based modelling and simulation of liquid immersion cooling systems},
  author  = {Lionello, Michele and Rampazzo, Mirco and Beghi, Alessandro and Varagnolo, Damiano and Vesterlund, Mattias},
  journal = {Energy},
  volume  = {207},
  number  = {},
  pages   = {},
  month   = sep,
  year    = {2020},
  doi     = {10.1016/j.energy.2020.118238}
}

@article{Yang_2018ATE,
  title   = {Modeling, cross-validation, and optimization of a shipboard integrated energy system cooling network},
  author  = {Yang, S and Chagas, M B and Ordonez, J C},
  journal = {Applied Thermal Engineering},
  volume  = {145},
  number  = {},
  pages   = {516--527},
  month   = dec,
  year    = {2018},
  doi     = {10.1016/j.applthermaleng.2018.09.070}
}

@article{Koeln_202CST,
  title   = {Hierarchical Control of Aircraft Electro-Thermal Systems},
  author  = {Koeln, J P and Pangborn, H C and Williams, M A and Kawamura, M L and Alleyne, A G},
  journal = {IEEE Transactions on Control Systems Technology},
  volume  = {28},
  number  = {4},
  pages   = {1218--1232},
  month   = jul,
  year    = {2020},
  doi     = {10.1109/TCST.2019.2905221}
}

@article{Pangborn2017,
  title   = {Hardware-in-the-Loop Validation of Advanced Fuel Thermal Management Control},
  author  = {Pangborn, Herschel C and Hey, Joel E and Deppen, Timothy O and Alleyne, Andrew G and Fisher, Timothy S},
  journal = {Journal of Thermophysics and Heat Transfer},
  volume  = {31},
  number  = {4},
  pages   = {901--909},
  month   = oct,
  year    = {2017},
  doi     = {10.2514/1.T5055}
}

@article{Laird_2022Energy,
  title   = {Framework for integrated plant and control optimization of electro-thermal systems: An energy storage system case study},
  author  = {Laird, Cary and Kang, Ziliang and James, Kai A and Alleyne, Andrew G},
  journal = {Energy},
  volume  = {258},
  number  = {1},
  pages   = {},
  month   = nov,
  year    = {2022},
  doi     = {10.1016/j.energy.2022.124855}
}

@article{jahan2024study,
  title   = {A Study on Control Co-Design for Optimizing Microgrid Sustainability},
  author  = {Jahan, Tania Rifat and Ouedraogo, Asmaou S and Docimo, Donald J},
  journal = {{IFAC}-{PapersOnLine}},
  volume  = {58},
  number  = {28},
  pages   = {636--641},
  year    = {2024},
  doi     = {10.1016/j.ifacol.2025.01.037}
}

@article{Docimo_2021JDMC,
  title   = {Plant and controller optimization for power and energy systems with model predictive control},
  author  = {Docimo, Donald J and Kang, Ziliang and James, Kai A and Alleyne, Andrew G},
  journal = {Journal of Dynamic Systems, Measurement, and Control},
  volume  = {143},
  number  = {8},
  pages   = {},
  month   = aug,
  year    = {2021},
  doi     = {10.1115/1.4050399}
}

@article{koot2021usage,
  title   = {Usage impact on data center electricity needs: A system dynamic forecasting model},
  author  = {Koot, Martijn and Wijnhoven, Fons},
  journal = {Applied Energy},
  volume  = {291},
  number  = {},
  pages   = {},
  month   = jun,
  year    = {2021},
  doi     = {10.1016/j.apenergy.2021.116798}
}

@article{Aksland_2023CST,
  title   = {Hierarchical Predictive Control of an Unmanned Aerial Vehicle Integrated Power, Propulsion, and Thermal Management System},
  author  = {Aksland, C T and Tannous, P J and Wagenmaker, M J and Pangborn, H C and Alleyne, A G},
  journal = {IEEE Transactions on Control Systems Technology},
  volume  = {31},
  number  = {3},
  pages   = {1280--1295},
  month   = may,
  year    = {2023},
  doi     = {10.1109/TCST.2022.3220913}
}

@article{ahu2026,
    title   = {Control-oriented design framework for heat pump-based thermal management systems of electric vehicles},
    author  = {Ahn, Changsun and Sun, Jing},
    journal = {Control Engineering Practice},
    volume  = {172},
    number  = {},
    page    = {},
    month   = jul,
    year    = {2026},
    doi     = {10.1016/j.conengprac.2026.106946},
}

@article{Tannous2020_JDMC,
    title   = {Multilevel Hierarchical Estimation for Thermal Management Systems of Electrified Vehicles With Experimental Validation},
    author  = {Tannous, Pamela J. and Alleyne, Andrew G.},
    journal = {Journal of Dynamic Systems, Measurement, and Control},
    volume  = {142},
    number  = {},
    page    = {},
    month   = nov,
    year    = {2020},
    doi     = {10.1115/1.4047786},
}

@article{Peddada2020_JDM,
    title   = {Optimal Sensor Placement Methods in Active High Power Density Electronic Systems With Experimental Validation},
    author  = {Peddada, Satya R. T. and Tannous, Pamela J. and Alleyne, Andrew G. and Allison, James T.},
    journal = {Journal of Mechanical Design },
    volume  = {142},
    number  = {},
    page    = {},
    month   = feb,
    year    = {2020},
    doi     = {10.1115/1.4044744},
}

@article{Tannous2019_CEP,
    title   = {Model-based temperature estimation of power electronics systems},
    author  = {Tannous, Pamela J. and Peddada, Satya R.T. and Allison, James T. and Foulkesc, Thomas and Pilawa-Podgurskic, Robert C.N. and Alleyne, Andrew G.},
    journal = {Control Engineering Practice},
    volume  = {85},
    number  = {},
    page    = {},
    month   = apr,
    year    = {2019},
    doi     = {10.1016/j.conengprac.2019.01.006},
}

@article{Tannous2019_JDMC,
    title   = {Fault Detection and Isolation for Complex Thermal Management Systems},
    author  = {Tannous, Pamela J. and Alleyne, Andrew G.},
    journal = {Journal of Dynamic Systems, Measurement, and Control},
    volume  = {141},
    number  = {},
    page    = {},
    month   = jun,
    year    = {2019},
    doi     = {10.1115/1.4042675},
}

@inproceedings{Tannous2019_ACC,
  title     = {Hierarchical Estimation for Complex Multi-Domain Dynamical Systems}, 
  author    = {Tannous, Pamela J. and Docimo, Donald J. and Pangborn, Herschel C. and Alleyne, Andrew G.},
  booktitle = {2019 Annual American Control Conference (ACC)}, 
  month     = jul,
  year      = {2019},
  eventdate = {2019-07-10/2019-07-12},
  address   = {Philadelphia, PA, USA},
  pages     = {},
  doi       = {10.23919/ACC.2019.8814330}
  }

@inproceedings{Tannous2017_ACC,
  title     = {Dynamic temperature estimation of power electronics systems}, 
  author    = {Tannous, Pamela J. and Peddada, Satya R. T. and Allison, James T. and Foulkes, Thomas and Pilawa-Podgurski, Robert C.N. and Alleyne, Andrew G.},
  booktitle = {2017 American Control Conference (ACC)}, 
  month     = may,
  year      = {2017},
  eventdate = {2017-05-24/2017-05-26},
  address   = {Seattle, WA, USA},
  pages     = {},
  doi       = {10.23919/ACC.2017.7963482}
  }

@inproceedings{Peddada2017DETC,
  title     = {OPTIMAL SENSOR PLACEMENT METHODS FOR ACTIVE POWER ELECTRONIC SYSTEMS}, 
  author    = {Peddada, Satya R. T. and Tannous, Pamela J. and Alleyne, Andrew G. and Allison, James T.},
  booktitle = {ASME 2017 International Design Engineering Technical Conferences and Computers and Information in Engineering Conference}, 
  month     = aug,
  year      = {2017},
  eventdate = {2017-08-06/2017-08-09},
  address   = {Cleveland, Ohio, USA},
  pages     = {},
  doi       = {}
  }

@inproceedings{Tannous2020,
  title     = {MULTI-LEVEL HIERARCHICAL ESTIMATION FOR THERMAL MANAGEMENT SYSTEMS OF ELECTRIFIED VEHICLES}, 
  author    = {Tannous, Pamela J. and Alleyne, Andrew G.},
  booktitle = {ASME 2020 Dynamic Systems and Control Conference}, 
  month     = oct,
  year      = {2020},
  eventdate = {2020-10-05/2020-10-07},
  address   = {Virtual, Online},
  pages     = {},
  doi       = {}
  }

@book{Wen2018,
editor = {Wen, John T and Mishra, Sandipan},
isbn = {9783319684611},
publisher = {Springer},
title = {{Intelligent Building Control Systems}},
year = {2018}
}

@book{khalil_nonlinear_2002,
  address = {Upper Saddle River, {N.J.}},
  title = {Nonlinear systems},
  isbn = {0130673897 9780130673893 0131227408 9780131227408},
  language = {English},
  publisher = {Prentice Hall},
  author = {Khalil, Hassan K},
  year = {2002}
}

@inproceedings{Bolander_2025SCITECH,
    author = {Ara Bolander and Trevor Bird and William A. Malatesta and Kevin McCarthy and Neera Jain},
    title = {A Multi-state Graph-based Framework for Dynamic Modeling of Turbomachinery Components},
    booktitle = {{AIAA} {SCITECH} 2024 Forum},
    address = {Orlando, FL, USA},
    month = jan,
    year = 2024,
    doi  = {10.2514/6.2024-0162}
}

@inproceedings{Pangborn_2018ACC,
  title     = {Passivity and Decentralized {MPC} of Switched Graph-Based Power Flow Systems}, 
  author    = {Pangborn, Herschel C. and Koeln, Justin P. and Alleyne, Andrew G.},
  booktitle = {2018 Annual American Control Conference (ACC)}, 
  month     = jun,
  year      = {2018},
  eventdate = {2018-06-27/2018-06-29},
  address   = {},
  pages     = {198-203},
  doi       = {10.23919/ACC.2018.8431722}
  }

@inproceedings{PangbornACC2019,
  title     = {Cooperativity and Hierarchical {MPC} of State-Constrained Switched Power Flow Systems}, 
  author    = {Pangborn, Herschel C. and Alleyne, Andrew G.},
  booktitle = {2019 American Control Conference (ACC)}, 
  month     = jul,
  year      = {2019},
  address   = {Philadelphia, PA, USA},
  pages     = {4245-4252},
  doi       = {10.23919/ACC.2019.8815363}
  }

@inproceedings{Smith_2025ACC,
   title     = {Decomposition-Based Control Co-Design of Energy Systems Using Graph Models}, 
   author    = {Smith, Kayla Russell and Alleyne, Andrew G.},
   booktitle = {2025 American Control Conference (ACC)}, 
   month     = jul,
   year      = {2025},
   address   = {Denver, CO, USA},
   pages     = {4743-4749},
   doi       = {10.23919/ACC63710.2025.11107687}
  }

@inproceedings{Smith_2022ACC,
   title     = {Dynamical Graph-Based Models of Brayton Cycle Systems}, 
   author    = {Smith, Reid D. and Alleyne, Andrew G.}, 
   booktitle = {2022 American Control Conference (ACC)}, 
   month     = jun,
   year      = {2022},
   address   = {Atlanta, GA, USA},
   pages     = {4802-4807},
   doi       = {10.23919/ACC53348.2022.9867330}
}

@inproceedings{Thompson_2023CCTA,
   title     = {Combined Design and Open-loop Control Optimization for Propulsion, Power, and Thermal Management of Hybrid-electric Aircraft}, 
   author    = {Thompson, Andrew F. and Iezzi, Andrew J. and Pangborn, Herschel C. and Madabhushi, Ravi K. and Atassi, Oliver V.},
   booktitle = {2023 IEEE Conference on Control Technology and Applications (CCTA)}, 
   month     = aug,
   year      = {2023},
   address   = {Bridgetown, Barbados},
   pages     = {955-962},
   doi       = {10.1109/CCTA54093.2023.10252309}
}

@inproceedings{Aksland_2019ACC,
   title     = {Graph-Based Electro-Mechanical Modeling of a Hybrid Unmanned Aerial Vehicle for Real-Time Applications}, 
   author    = {Aksland, Christopher T and Bixel, Tyler W and Raymond, Logan C and Rottmayer, Michael A and Alleyne, Andrew G},
   booktitle = {2019 American Control Conference (ACC)}, 
   month     = jul,
   year      = {2019},
   address   = {Philadelphia, PA, USA},
   pages     = {4253--4259},
   doi       = {10.23919/ACC.2019.8814930}
}

@inproceedings{Yu_2023Aviation,
   title     = {Graph-based hierarchical control of electrified aircraft systems with automated timescale decomposition}, 
   author    = {Yu, Yin and Park, Seho and Huang, Daning and Pangborn, Herschel},
   booktitle = {AIAA AVIATION 2023 Forum}, 
   month     = jun,
   year      = {2023},
   address   = {San Diego, CA, USA},
   pages     = {},
   doi       = {10.2514/6.2023-4509}
}

@inproceedings{Manion_2022iTherm,
   title     = {Development of a Graph-based Modeling Framework for Transient Exergy Analysis}, 
   author    = {Manion, A R and Malatesta, W A and Jain, N},
   booktitle = {2022 21st IEEE Intersociety Conference on Thermal and Thermomechanical Phenomena in Electronic Systems (iTherm)}, 
   month     = May,
   year      = {2022},
   address   = {San Diego, CA, USA},
   pages     = {1--10},
   doi       = {10.1109/iTherm54085.2022.9899658}
}

@inproceedings{Park_2023CCTA,
   title     = {Lifted Graph-based Modeling for Linear Predictive Control of Nonlinear Energy Systems}, 
   author    = {Park, Seho and Pangborn, Herschel C.},
   booktitle = {IEEE Conference on Control Technology and Applications (CCTA)}, 
   month     = aug,
   year      = {2023},
   address   = {Bridgetown, Barbados},
   pages     = {926--933},
   doi       = {10.1109/CCTA54093.2023.10252310},
}

@inproceedings{Koeln2015DSCC,
   title     = {Hierarchical Control of Multi-Domain Power Flow in Mobile Systems: Part {I} — Framework Development and Demonstration},
   author    = {Koeln, Justin P and Williams, Matthew A and Alleyne, Andrew G},
   booktitle = {Dynamic Systems and Control Conference},
   month     = oct,
   year      = {2015},
   address   = {Columbus, OH, USA},
   pages     = {},
   doi       = {10.1115/DSCC2015-9908}
}

@inproceedings{Docimo_2018CCTA,
   title     = {Electro-Thermal Graph-Based Modeling for Hierarchical Control with Application to an Electric Vehicle},
   author    = {Docimo, Donald J and Alleyne, Andrew G},
   booktitle = {IEEE Conference on Control Technology and Applications (CCTA)},
   month     = aug,
   year      = {2018},
   address   = {Copenhagen, Denmark},
   pages     = {812--819},
   doi       = {10.1109/CCTA.2018.8511390}
}

@inproceedings{Williams2015DSCC,
   title     = {Hierarchical Control of Multi-Domain Power Flow in Mobile Systems: Part {II} — Aircraft Application},
   author    = {Williams, Matthew A and Koeln, Justin P and Alleyne, Andrew G},
   booktitle = {Dynamic Systems and Control Conference},
   month     = oct,
   year      = {2015},
   address   = {Columbus, OH, USA},
   pages     = {},
   doi       = {10.1115/DSCC2015-9904}
}

@inproceedings{Kolen_2016DSCC,
   title     = {Experimental Validation of Graph-Based Modeling for Thermal Fluid Power Flow Systems}, 
   author    = {Koeln, Justin P and Williams, Matthew A and Pangborn, Herschel C and Alleyne, Andrew G},
   booktitle = {Dynamic Systems and Control Conference}, 
   month     = oct,
   year      = {2016},
   address   = {Minneapolis, Minnesota, USA},
   pages     = {},
   doi       = {10.1115/DSCC2016-9782}
}

@inproceedings{Docimo2022ACC,
   title     = {A design framework with embedded hierarchical control architecture optimization}, 
   author    = {Docimo, Donald J},
   booktitle = {2022 American Control Conference (ACC)},
   month     = jun,
   year      = {2022},
   address   = {Atlanta, GA, USA},
   pages     = {3184--3191},
   doi       = {10.23919/ACC53348.2022.9867555}
}

@inproceedings{Aksland_2017DSCC,
   title     = {A graph-based approach for dynamic compressor modeling in vapor compression systems}, 
   author    = {Aksland, Christopher T. and Koeln, Justin P. and Alleyne, Andrew G.},
   booktitle = {{ASME} 2017 Dynamic Systems and Control Conference}, 
   month     = oct,
   year      = {2017},
   address   = {Tysons, VA, USA},
   pages     = {},
   doi       = {10.1115/DSCC2017-5238},
}

@inproceedings{Pangborn_2017ACC,
   title     = {Graph-based hierarchical control of thermal-fluid power flow systems}, 
   author    = {Pangborn, Herschel C. and Williams, Matthew A. and Koeln, Justin P. and Alleyne, Andrew G.},
   booktitle = {2017 American Control Conference ({ACC})}, 
   month     = may,
   year      = {2017},
   address   = {Seattle, WA, USA},
   pages     = {2099-2105},
   doi       = {10.23919/ACC.2017.7963262}
}

@inproceedings{Vyas_2026SCITECH,
   title     = {Graph-based dynamic modeling of a {PAO} loop for aircraft thermal management systems}, 
   author    = {Vyas, Vismay and Jain, Neera and McCarthy, Kevin and Bolander, Ara},
   booktitle = {{AIAA} {SCITECH} 2026 Forum}, 
   month     = jan,
   year      = {2026},
   address   = {Orlando, FL, USA},
   pages     = {},
   doi       = {10.2514/6.2026-2460}
}

@inproceedings{Echreshavi_2025CLIMA,
   title     = {Control-oriented graph-based modelling of building energy system: a conservation-based framework for multi-zone buildings}, 
   author    = {Echreshavi, Zeinab and Sisti, Enrico and Palangari, Mohsen Farbood and Carli, Ruggero and Rampazzo, Mirco},
   booktitle = {5th {REHVA} {HVAC} World Congress - {CLIMA} 2025}, 
   month     = jun,
   year      = {2025},
   address   = {Milan, Italy},
   pages     = {137-147},
   doi       = {0.1007/978-3-032-06806-4_14}
}

@inproceedings{Sisti_2025CLIMA,
   title     = {Modeling and simulation of vapor compression systems using a graph-based toolbox}, 
   author    = {Sisti, Enrico and Rampazzo, Mirco and Beghi, Alessandro},
   booktitle = {Proceedings of the 15th {REHVA} {HVAC} World Congress - {CLIMA} 2025}, 
   month     = jun,
   year      = {2025},
   address   = {Milan, Italy},
   pages     = {1008-1017},
   doi       = {10.1007/978-3-032-06806-4_97}
}

@inproceedings{Pangborn_2020ACC,
   title     = {Hierarchical hybrid {MPC} for management of distributed phase change thermal energy storage}, 
   author    = {Pangborn, Herschel C. and Laird, Cary E. and Alleyne, Andrew G.},
   booktitle = {2020 American Control Conference ({ACC})}, 
   month     = jul,
   year      = {2020},
   address   = {Denver, CO, USA},
   pages     = {4147-4153},
   doi       = {10.23919/ACC45564.2020.9147698}
}

@inproceedings{Belkacem_2025FKFS,
   title     = {Topology and sizing optimization of thermal and electric energy systems for battery electric vehicle}, 
   author    = {Belkacem, Fakher and Schäfer, Henrik and Hellberg, Tobias and Meywerk, Martin},
   booktitle = {2025 {FKFS} Conference on Vehicle Aerodynamics and Thermal Management}, 
   month     = oct,
   year      = {2025},
   address   = {Leinfelden-Echterdingen, Germany},
   pages     = {},
   doi       = {}
}

@inproceedings{Laird_2020DSCC,
   title     = {Graph-based design and control optimization of a hybrid electrical energy storage system}, 
   author    = {Laird, Cary and Docimo, Donald and Aksland, Christopher T. and Alleyne, Andrew G.},
   booktitle = {{ASME} 2020 Dynamic Systems and Control Conference}, 
   month     = oct,
   year      = {2020},
   address   = {Virtual, Online},
   pages     = {},
   doi       = {10.1115/DSCC2020-3233}
}

@inproceedings{Walters2010,
   title     = {{INVENT} Modeling, Simulation, Analysis and Optimization}, 
   author    = {Walters, Eric A. and Iden, Steve},
   booktitle = {48th AIAA Aerospace Sciences Meeting}, 
   month     = jan,
   year      = {2010},
   address   = {Orlando, Florida, USA},
   pages     = {1--11},
   doi       = {10.2514/6.2010-287}
}

@inproceedings{kolen2025ACC,
  title   = {Improved Mass Conservation of Control-Oriented Models of Two-Phase Thermal Systems using Neural Networks},
  author  = {Gomez, Alexander M. and Shaikh, Juned and Koeln, Justin P.},
  booktitle = {2025 American Control Conference (ACC)},
  month   = jul,
  year    = {2025},
  address   = {Denver, CO, USA},
  pages   = {},
  doi     = {10.23919/ACC63710.2025.11107885}
}

@phdthesis{Pangborn2019,
author = {Pangborn, Herschel C.},
school = {University of Illinois at Urbana-Champaign},
title = {{Hierarchical Control for Multi-Domain Coordination of Vehicle Energy Systems with Switched Dynamics}},
year = {2019}
}

@phdthesis{Tannous2019,
author = {Tannous, Pamela J.},
school = {University of Illinois at Urbana-Champaign},
title = {{Estimation and Fault Diagnosis for Vehicle Energy Systems}},
year = {2020}
}

@article{yu2024learningnetworkeddynamicalmodels,
      title={Learning Networked Dynamical System Models with Weak Form and Graph Neural Networks}, 
      author={Yin Yu and Daning Huang and Seho Park and Herschel C. Pangborn},
      year={2024},
      journal={arXiv 2407.16779},
      eprint={2407.16779},
      archivePrefix={arXiv},
      primaryClass={eess.SY},
      url={https://arxiv.org/abs/2407.16779}, 
}

@misc{MATLAB,
year = {2025},
author = {{The MathWorks Inc.}},
title = {{MATLAB} version: 25.1.0 (2025a)},
publisher = {The MathWorks Inc.},
address = {Natick, Massachusetts, United States},
url = {https://www.mathworks.com}
}
\bibliographystyle{ieeetr}

\end{document}